\documentclass[pra,twocolumn,superscriptaddress,floatfix]{revtex4}

\usepackage{amsmath}
\usepackage{amstext}
\usepackage{amssymb}
\usepackage{bbm}
\usepackage{latexsym}
\usepackage[dvips]{graphicx}

\newcommand{\beq}{\begin{equation}}
\newcommand{\eeq}{\end{equation}}
\newcommand{\barr}{\begin{eqnarray}}
\newcommand{\earr}{\end{eqnarray}}
\newcommand{\ket}[1]{\left\vert#1\right\rangle}
\newcommand{\bra}[1]{\left\langle#1\right\vert}
\newcommand{\one}{\mathbbm{1}}
\newcommand{\Ham}{\mathcal H}

\begin{document}
\title{Decoherence induced by interacting quantum spin baths}

\author{Davide Rossini}
\affiliation{NEST-CNR-INFM \& Scuola Normale Superiore,
  Piazza dei Cavalieri 7, I-56126 Pisa, Italy}
\author{Tommaso Calarco}
\affiliation{Dipartimento di Fisica,
  Universit\`a di Trento and BEC-CNR-INFM, I-38050 Povo, Italy}
\affiliation{ITAMP, Harvard-Smithsonian Center for Astrophysics,
  Cambridge, MA 02138, USA}
\author{Vittorio Giovannetti}
\affiliation{NEST-CNR-INFM \& Scuola Normale Superiore,
  Piazza dei Cavalieri 7, I-56126 Pisa, Italy}
\author{Simone Montangero}
\affiliation{NEST-CNR-INFM \& Scuola Normale Superiore,
  Piazza dei Cavalieri 7, I-56126 Pisa, Italy}
\author{Rosario Fazio}
\affiliation{NEST-CNR-INFM \& Scuola Normale Superiore,
  Piazza dei Cavalieri 7, I-56126 Pisa, Italy}
\affiliation{International School for Advanced Studies (SISSA),
  Via Beirut 2-4, I-34014 Trieste, Italy}

\date{\today}

\begin{abstract}

  We study decoherence induced on a two-level system coupled
  to a one-dimensional quantum spin chain.
  We consider the cases where the dynamics of the chain is 
  determined by the Ising, $XY$, or Heisenberg exchange Hamiltonian.
  This model of quantum baths can be of fundamental importance for the
  understanding of decoherence in open quantum systems, since it
  can be experimentally engineered by using atoms in optical lattices.
  As an example, here we show how to implement a pure dephasing model
  for a qubit system coupled to an interacting spin bath.
  We provide results that go beyond the case of a central spin
  coupled uniformly to all the spins of the bath, in particular showing
  what happens when the bath enters different phases, or becomes critical;
  we also study the dependence of the coherence loss on the number
  of bath spins to which the system is coupled and we describe a
  coupling-independent regime in which decoherence exhibits universal
  features, irrespective of the system-environment coupling strength.
  Finally, we establish a relation between decoherence and entanglement
  inside the bath.
  For the Ising and the $XY$ models we are able to give an exact
  expression for the decay of coherences, while for the Heisenberg
  bath we resort to the numerical time-dependent
  Density Matrix Renormalization Group.

\end{abstract}


\maketitle

\section{Introduction} 
\label{sec:intro}

Decoherence refers to the process through which superpositions of quantum
states are irreversibly transformed into statistical mixtures.
It is due to the unavoidable coupling of a quantum system with
its surrounding environment which, as a consequence, leads to entanglement
between the system and the bath. 
This loss of coherence may considerably reduce the efficiency of quantum 
information protocols~\cite{nielsen} and it is crucial in describing
the emergence of classicality in quantum systems~\cite{zurek}. 

Although desirable, it is not always possible to fully characterize
the bath. Therefore it is necessary to resort to ingenious,
though realistic, modelizations. Paradigmatic models represent the environment
as a many-body system, such as a set of bosonic harmonic
oscillators~\cite{caldeira,weiss} or of spin-1/2 particles~\cite{prokofiev}.
In some cases it is also possible to recover the effects
of a many-body environment via the coupling to a single-particle
bath, provided its dynamics is chaotic~\cite{benenti}.
In order to grasp all the subtleties of the entanglement between
the system and its environment, it would be of great importance to study
engineered baths (and system-bath interactions) that can be realized
experimentally.
In this paper we discuss a class of {\em interacting spin baths} which
satisfy these requirements: a two-level system coupled to a one-dimensional
array of spin-1/2 particles, whose free evolution is driven by 
a Hamiltonian which embraces Ising, $XY$ and Heisenberg universality classes.
In several non-trivial cases we can solve the problem {\em exactly}.
Moreover we show that it is possible to realize these system+bath Hamiltonians 
with cold bosons in optical lattices. We thus extend the approach of 
Jan\'e {\em et al.}~\cite{jane03} showing that optical lattices can be useful
as {\em open quantum system simulators}.

Our analysis can be framed in the context of the recently growing interest 
in the study of decoherence due to spin baths, see~\cite{zurek2,cucchietti,
tessieri,dawson,paganelli,quan,cucchietti2,ou06,khveshchenko,dobrovitski,lages,camalet,zurek3}. 
Starting from the seminal paper of Zurek~\cite{zurek2}, several papers
analyzed the decoherence due to a collection of independent spins.
Cucchietti, Paz and Zurek~\cite{cucchietti} derived several properties
of the Loschmidt echo starting from fairly general assumptions for
the distribution of the splittings of the bath spins.
The effect of an infinite-range interaction among the spins was introduced
in Ref.~\cite{tessieri}; the same model was further exploited
in Dawson {\em et al.}~\cite{dawson} to relate decoherence to the entanglement
present in the bath. Properties of decoherence in presence
of symmetry breaking, and the effect of critical behavior of the bath were
discussed in~\cite{paganelli,quan}.   
Most of the works done so far are based on the so called central spin model,
where the two-level system is coupled isotropically to all the spins
of the bath. This assumption tremendously simplifies the derivation,
but at the same time it may introduce some fictitious symmetries
which are absent in realistic systems.
Moreover it can be very hard to simulate it with engineered baths.
A crucial feature of our work is the assumption that the two-level system 
interacts with only {\em few} spins of the bath. As we will show,
this introduces qualitative differences as compared to the central spin model,
moreover it is amenable to an experimental implementation
with optical lattices. 

The paper is organized as follows. In the next Section we introduce
the system+bath model Hamiltonians that will be studied throughout this paper
and we define the central quantity that will be used in order to characterize
the decoherence of the system, i.e. the Loschmidt echo. 
We then show in Section~\ref{sec:simulation} 
how optical lattices can be used to simulate the class of Hamiltonians
introduced in Section~\ref{sec:model}.
The rest of the paper is devoted to the derivation and to the 
analysis of our results, both for a single system-bath link
(Section~\ref{sec:results}), and for multiple links
(Section~\ref{sec:manylinks}).
When the two-level system is coupled to an $XY$- or $XX$-model, it is possible 
to derive an exact result for the Loschmidt echo.
This is explained in Section~\ref{subsec:XYanalyt}, where we also discuss
in detail its short- and long-time behavior and relate it to 
the critical properties of the chain. Further insight is obtained
by perturbative calculations which agree very well, in the appropriate limits,
with the exact results. In Section~\ref{subsec:Heisenberg} we present
our results for the Heisenberg bath.
In this case an analytic approach is not possible.
Here we solve the problem by means of the time-dependent Density 
Matrix Renormalization Group (t-DMRG -- see Appendix~\ref{DMRG}). 
In Section~\ref{sec:deco_ent} we analyze the possible relation between
decoherence and entanglement properties of the environment:
we relate the short-time decay of coherences to the two-site
nearest-neighbor concurrence inside the bath.
In Section~\ref{sec:manylinks} we extend our results to the case
in which the system is coupled to an arbitrary number of bath spins;
a regime in which the decoherence is substantially independent of the
coupling strength between the system and the environment is discussed in
Section~\ref{sec:strongpert}.
Finally, in Section~\ref{sec:conclusions} we draw our conclusions.
In the Appendices we give some technical details on the numerical DMRG
approach (Appendix~\ref{DMRG}), we provide explicit expressions for the
Fermion correlation functions needed for the evaluation of the Loschmidt echo
in the $XY$-model (Appendix~\ref{fermion}) and we briefly discuss the
central spin model (Appendix~\ref{central}) to make our paper self-contained.
A brief account of some of the results discussed in this paper
is presented in~\cite{rossini06}.

\section{The model} 
\label{sec:model}

\begin{figure}[!ht]
  \begin{center}
    \includegraphics[scale=0.43]{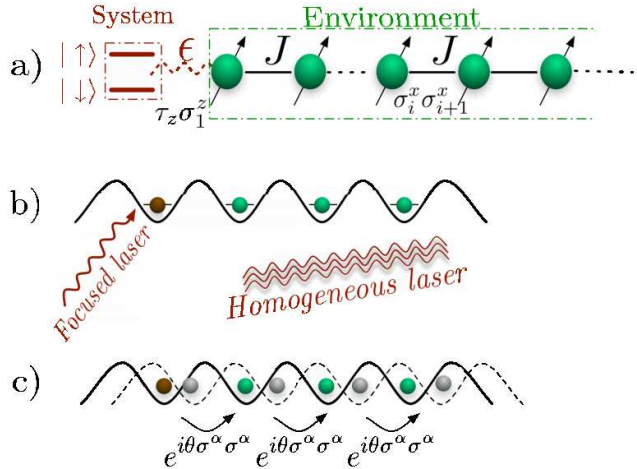}
    \caption{a) A sketch of the system-plus-bath model we consider in this
    work. The two-level system (at position zero) is coupled to the
    $\sigma^z$ component of the first spin of the chain that acts as
    a spin bath. Atoms in an optical lattice can simulate
    this controlled decoherence by means of series of lasers b) and
    lattice displacements c), which allow to realize both the
    interaction of the bath with an external magnetic field and the
    anisotropic exchange coupling present in Eq.~(\ref{eq:spinbath}).}
    \label{fig:scheme}
  \end{center}
\end{figure}

The model we consider consists of a two-level quantum object (qubit) $S$ 
coupled to an {\em interacting} spin bath $E$ composed by $N$ spin-$1/2$ 
particles (see Fig.\ref{fig:scheme}a): 
the idea is to study how the internal dynamics of $E$ affects the 
decoherent evolution of $S$.

In our scheme the global system $S+E$ is fully characterized
by a standard Hamiltonian of the form
\beq
     \Ham = \Ham_S + \Ham_E + \Ham_{\mathrm{int}} \, ,
     \label{eq:system+bath}
\eeq
with $\Ham_{S,E}$ being the free-Hamiltonians of $S$ and $E$, and 
$\Ham_{\mathrm{int}}$ being the coupling term. Without loss of generality
we will assume the free Hamiltonian of the qubit to be of the form 
\beq
     \label{eq:free}
     \Ham_S = \frac{\omega_e}{2} \left( \one - \tau_z \right)
     = \omega_e \ket{e} \bra{e} \, ,
\eeq
with $\tau_\alpha$ being the Pauli matrices of $S$ ($\alpha = x,y,z$),
and $\ket {e}$ its excited state (the ground state being
represented by the vector $\ket {g}$).
On the other hand, the environment will be modeled by a one-dimensional
quantum spin-1/2 chain described by the Hamiltonian 
\barr 
     \label{eq:spinbath}
     \Ham_E & = & -\frac{J}{2} \sum_j \left[
     \left( 1 + \gamma \right) \sigma^x_j \sigma^x_{j+1} +
     \left( 1 - \gamma \right) \sigma^y_j \sigma^y_{j+1} \right. \nonumber \\
     & & + \left. \Delta \sigma^z_j \sigma^z_{j+1} +
     2 \lambda \sigma^z_j \right] \, ,
\earr
where $\sigma^\alpha_i$ ($\alpha = x,y,z$) are the Pauli matrices
of the $i$-th spin.
The sum over $j$ goes from 1 to $N-1$ for open boundary conditions,
or from 1 to $N$ for periodic boundary conditions (where we assume that
$\sigma^\alpha_{N+1} \equiv \sigma^\alpha_{1}$).
The constants $J$, $\Delta$, $\gamma$ and $\lambda$ respectively characterize
the interaction strength between neighboring spins, the anisotropy
parameter along $z$ and in the $xy$ plane,
and an external transverse magnetic field.
The Hamiltonian~\eqref{eq:spinbath} has
a very rich structure~\cite{sachdev99}.  For the sake of
simplicity, we shall consider 
the following paradigmatic cases:
\begin{itemize}
\item the $XY$-model in a transverse field -- see Sec.~\ref{subsec:XYanalyt}.
Here one has $\Delta =0$ and  $\lambda$, $\gamma$ generic.
For $0 < \gamma \leq 1$, Eq.~\eqref{eq:spinbath} belongs to the 
Ising universality class, and it has a critical point at
$\vert \lambda_c \vert = 1$; for $\gamma = 0$ it reduces to the
$XX$ universality class, which is critical for $\vert \lambda \vert \leq 1$.

\item the $XXZ$ anisotropic Heisenberg model -- see
Sec.~\ref{subsec:Heisenberg}.
Here one has $\lambda, \gamma=0$ and $\Delta$ generic. 
In this case the Hamiltonian~\eqref{eq:spinbath} is critical for
$-1 \leq \Delta \leq 1$ while it has ferromagnetic or 
anti-ferromagnetic order for $\Delta>1$ or $\Delta<-1$ respectively. 
\end{itemize}
Finally, the qubit $S$ is coupled to the spin bath through 
a dephasing interaction of the form 
\beq
     \Ham_{\mathrm{int}} =
     -\epsilon \sum_{j=j_1}^{j_m} \ket{e} \bra{e} \sigma_j^z \, ,
     \label{eq:hamint}
\eeq
where $\epsilon$ is the coupling constant, 
and the {\em link number} $m$ counts the number of
environmental spins (labeled by $j_1 \ldots j_m$) to which $S$ is coupled 
--- Fig.\ref{fig:scheme}a 
refers to the case where $S$ is interacting with the first spin
of an open-boundary chain, i.e. $m=1$, $j_1=1$.

By varying the parameters $m, \Delta, \gamma, \lambda$ and $\epsilon$, 
the above Hamiltonians allow us to analyze several non-trivial $S+E$ scenarios.
Moreover we will see in  Sec.~\ref{sec:simulation} that it is possible
to use optical lattices manipulation techniques~\cite{duan03,jane03}
to experimentally simulate the resulting dynamical evolution.

\subsection{Decoherence: the Loschmidt echo} 
\label{sec:echo}

With the choice of the coupling~(\ref{eq:hamint})  the populations  
of the ground and excited states of the qubit do not evolve in time,
since $[\tau^z, \Ham] =0$.
Consequently, no dissipation takes place in the model and  
the qubit evolution is purely decoherent:   
the system $S$ looses its coherence without exchanging energy
with the bath~\cite{unruh,palma}.
To study this effect of pure phase decoherence, we suppose that
at the beginning the qubit is completely disentangled from
the environment --- namely, at time $t=0$ the global system
wave-function is given by
\beq
     \ket{\Psi (0)} = \vert \phi (0) \rangle_S
     \otimes \vert \varphi (0) \rangle_E \, ,
\eeq
where $\vert \phi (0) \rangle_S = c_g \ket{g} + c_e \ket{e}$ is
a generic superposition of the ground and the excited state of $S$
and  $\vert \varphi (0) \rangle_E \equiv \ket{G}_E$ is the ground
state of the environment Hamiltonian $\Ham_E$.
The global evolution of the composite system under the Hamiltonian
Eq.~\eqref{eq:system+bath} will then split into two terms:
one where $E$ evolves with the {\em unperturbed} Hamiltonian
$\Ham_g \equiv \Ham_E$ and the other where $E$ evolves
with the {\em perturbed} Hamiltonian
\begin{eqnarray}
\Ham_e \equiv \Ham_E + \langle e \vert \Ham_{\mathrm{int}} \vert e \rangle
\label{eq:PERTHAM}\;.
\end{eqnarray}
As a result, one has
\barr
     \ket {\Psi (0)} \to \ket{\Psi (t)} & = &
     c_g \ket{g} \otimes \ket {\varphi_g (t)}_E \\
     &  + &c_e e^{-i \omega_e t} \ket{e} \otimes \ket {\varphi_e (t)}_E 
     \nonumber
\earr
where the two branches of the environment are
$\ket{\varphi_g (t)}_E = e^{-i \Ham_g t} \ket{\varphi (0)}_E$ and
$\ket{\varphi_e (t)}_E = e^{-i \Ham_e t} \ket{\varphi (0)}_E$.
Therefore the evolution of the reduced density matrix
$\rho \equiv \mathrm{Tr}_E \ket{\Psi} \bra{\Psi}$
of the two-level system corresponds to a pure dephasing process.
In the basis of the eigenstates $\{ \ket{g}$, $\ket{e} \}$, the diagonal terms
$\rho_{gg}$ and $\rho_{ee}$ do not evolve in time.
Instead the off-diagonal terms will decay according to
$$\rho_{eg} (t) = \rho_{eg} (0) \, e^{-i \omega_e t} D(t)\;,$$ 
where 
\beq
     D(t) \equiv \langle \varphi_g (t) \vert \varphi_e (t) \rangle =
     \langle \varphi (0) \vert e^{i \Ham_g t} e^{-i \Ham_e t}
     \vert \varphi (0) \rangle \, 
     \label{eq:loschdef}
\eeq
is the decoherence factor. 
The decoherence of $S$ can then be characterized
by the so-called ``Loschmidt echo'', i.e. by the real quantity   
\beq
     \mathcal{L} (t) \equiv  \vert D(t) \vert^2 = 
     \vert \langle G \vert e^{-i (\Ham_E +
       \langle e \vert \Ham_{\mathrm{int}} \vert e \rangle ) t}
     \vert G \rangle \vert^2 \, ,
     \label{eq:loschmidt} 
\eeq
where in the equality term  we used the fact that 
$ \vert \varphi (0) \rangle = \vert G \rangle$
is the ground state of $\Ham_E$. 
On one hand, values of $\mathcal{L}(t)$ close to $1$ indicate 
a weak interaction between the environment and the qubit
(the case $\mathcal{L}(t)=1$ corresponds to total absence
of interaction). On the other hand, values of $\mathcal{L}(t)$ 
close to $0$ correspond instead  to a
strong suppression of the qubit coherence due to the interaction with $E$
(for $\mathcal{L}(t)=0$ the qubit is maximally entangled
with the environment, and its density matrix $\rho$ becomes diagonal). 
In the following  
we will analyze the time evolution of the Loschmidt echo to
study how the internal dynamics of the environment $E$ affects the
decoherence of the system $S$. We will start with the case when
$S$ is coupled to a single spin of the chain $E$ (i.e. $m=1$, 
see Sec.~\ref{sec:results}). The case with multiple links 
will be instead discussed in Sec.~\ref{sec:manylinks}.

For the sake of completeness, we finally notice that, since the
function~(\ref{eq:loschmidt}) measures the overlap between the time-evolved
states of the same initial configuration under two slightly different
Hamiltonians, one can use it as an indicator of the stability of motion.
This kind of analysis has been performed in Ref.~\cite{zanardi06}, where the
fidelity between the ground states of quadratic Fermi Hamiltonians
has been analyzed in connection with quantum phase transitions.

\section{Simulation of open quantum systems by optical lattices} 
\label{sec:simulation}

Before analyzing in detail the temporal evolution of the 
Loschmidt echo~(\ref{eq:loschmidt}), in this Section 
we present a method which would allow one to experimentally simulate
the dynamics  induced by the Hamiltonian of Eq.~(\ref{eq:system+bath})
in a realistic setup.

The $S+E$ system introduced in Sec.~\ref{sec:model}
can be seen as an ``inhomogeneous'' spin network with $N+1$ sites,
where one of the spins (say, the first) plays the role of the system
of interest $S$ while the remaining $N$ play the role of the environment $E$.
This immediately suggests the possibility of simulating the dynamical
evolution of such a system on optical lattices by employing the techniques 
recently developed in Refs.~\cite{duan03,jane03}.
An important aspect of our scheme however is the fact that we assume 
the coupling between the system $S$ and $E$ to be independent 
from the couplings among the $N$ spins which compose the environment.
Analogously the free Hamiltonian of $S$ is assumed to be different from  
the on-site terms of the free-Hamiltonian of $E$. 
On one hand this allows us to study different environment Hamiltonians
without affecting the coupling between $E$ and $S$. On the other
hand, this also allow us to analyze different $S+E$ coupling regimes
(e.g. strong, weak) without changing the internal dynamics of the bath.

In order to include these crucial ingredients we found more convenient
to follow the scheme developed by Jan\'{e} {\em et al.} in Ref.~\cite{jane03} 
(the optical lattice manipulation schemes developed by Duan {\em et al.} 
in Ref.~\cite{duan03} seem to be less adequate to our purposes).
 The key advantage is that
we can realize the system-plus-bath setup by using a single one-dimensional
lattice in which the quantum system is placed on a given lattice site
(for example the first one, as in Fig.~\ref{fig:scheme}b).
The different Hamiltonians for the system and for the bath are realized by
specific pulse sequences which simulate the dynamics of the model.
The same holds for the coupling Hamiltonian of the two-level system
with the bath, which is different from the couplings within the bath.
In Fig.~\ref{fig:scheme} the leftmost atom simulates the two-level system,
the coupling to the second site is the interaction between the quantum system
and the environment, the rest of the chain is the interacting spin environment.

Jan\'e {\it et al.}~\cite{jane03} showed that atoms loaded in an optical
lattice can simulate the evolution of a generic spin Hamiltonian
in a stroboscopic way when subjected to appropriate
laser pulses, Fig.~\ref{fig:scheme}b, and 
controlled displacements, Fig.~\ref{fig:scheme}c, which allow to implement
the single-site and two-site contributions to the Hamiltonian.
The key point is that in our case the sequences of gates need to allow
for discriminating between the system and the bath.
The types of baths that one can simulate by these means embrace Ising,
$XY$ and Heisenberg exchange Hamiltonian. Therefore, by varying
the parameters of the optical lattice, it is possible to test the impact of
the different phases (critical, ferromagnetic, anti-ferromagnetic, etc$\dots$)
of the environment on the decoherence of the two-level system. 

Following the idea of a Universal Quantum Simulator described in~\cite{jane03},
the time evolution operator associated with ${\cal H}$ over a time $t$
can be simulated by decomposing it into a product of operators
acting on very short times $\tau\ll t$:
\barr
     e^{-i{\cal H}t} & = & \lim_{n\to\infty} \Bigg[ U^z_0 (\omega_e \tau) \,
       U^{zz}_{0,1} (\epsilon\tau) \, \prod_{j=1}^N
       U^z_j \big( \textstyle \frac{J\lambda\tau}{2} \big)\label{strobo}\\
       & \times & U^{xx}_{j,j+1} \big( \textstyle \frac{J(1+\gamma)\tau}{2} \big) \,
       U^{yy}_{j,j+1} \big( \textstyle \frac{J(1-\gamma)\tau}{2} \big) \,
       U^{zz}_{j,j+1} \big( \textstyle \frac{J\Delta\tau}{2} \big)
       \!\Bigg]^n \nonumber
\earr
where $\tau=t/n$, $U^z_j(\theta)\equiv e^{i\theta\sigma_j^z}$,
$U^{\alpha \beta}_{j, k}(\theta)\equiv e^{i\theta\sigma_j^\alpha\sigma_k^\beta}$,
and the index $0$ labels the two-level system $S$.
For $\alpha\in\{x,y\}$ one can write $U^{\alpha\alpha}_{j, k}=
V^\alpha_jV^\alpha_k U^{zz}_{j, k}V^{\alpha\dagger}_kV^{\alpha\dagger}_j$,
where $V^\alpha_j=(\openone-i\sigma^\alpha_j)/\sqrt{2}$ are
fast homogeneous local unitary operations. These can be realized with single
atoms trapped in an optical lattice~\cite{jane03}, each having two relevant
electronic levels ($|0\rangle_j \, , \,|1\rangle_{j}$) interacting
with a resonant laser according to:
\beq
     {\cal H}^L_j=\Omega \left( e^{i \phi} |1\rangle_j \langle0| +
     e^{-i\phi} |0\rangle_j \langle1| \, \right) \, .
     \label{eq:rabi}
\eeq

The evolution under the Hamiltonian Eq.~\eqref{eq:rabi},
$U^L_j(t,\phi)\equiv e^{-i{\cal H}^L_j t}$,
yields the single-qubit operations
$V^x_j = U^L_j ( \frac{\pi}{4\Omega}, 0)$,
$V^y_j = U^L_j ( \frac{\pi}{2\Omega}, 0) \, U^L_j ( \frac{3\pi}{4\Omega}, \frac{\pi}{2})$
and $U^z_j(\theta) = U^L_j( \frac{\pi}{2\Omega}, \pi+\theta) \, U^L_j( \frac{\pi}{2\Omega}, 0)$,
whence $U^z_j(-\frac\pi 2)=i\sigma^z_j$, while
$U^L_j(\frac\pi{2\Omega})=i\sigma^x_j$.
These operation can be made very fast by simply increasing the laser
intensity and thereby the Rabi frequency.
They can be performed either simultaneously on all qubits,
by shining the laser homogeneously onto all atoms, or selectively on some
of them, by focusing it appropriately (see Fig.~\ref{fig:scheme}b).
For our purposes the individual addressing is needed only for the atom
in position $0$, which represents the quantum system;
this is anyway the minimal physical requirement for being able to
monitor its state during the evolution.

Two-qubit operations can be performed by displacing the lattice
in a state-selective way~\cite{Jaksch}, 
so that state $|0\rangle_j|1\rangle_{j+1}$ acquires a phase factor
$e^{-i\varphi}$, as experimentally realized in~\cite{Bloch}. The resulting gate
$G_{j,j+1}(\varphi)$ can be composed with $\sigma^x$ rotations to yield 
$U^{zz}_{j,j+1} (\theta) = e^{i\theta} [G_{j,j+1} (2\theta) \, \sigma^x_j \sigma^x_{j+1}]^2$.
This will affect all atoms from $0$ to $N$.
Since we want a different coupling for the $\{01\}$ pair than for all others,
we need to erase the effect of the interaction for that specific pair
using only local operations on atom $0$, as in the sequences
$
[\sigma^z_0 \, U^{xx}_{01}(\theta)]^2 = [\sigma^z_0 \, U^{yy}_{01}(\theta)]^2 =
[\sigma^x_0 \, U^{zz}_{01}(\theta)]^2 = \openone.
$
Defining
$U^{\alpha\alpha}_\otimes ( \theta )\equiv \prod_{j=1}^N U^{\alpha\alpha}_{j,j+1}(\theta)$,
we can generate each simulation step in Eq.~(\ref{strobo}) as
\barr
    {\cal U}_n & \equiv &
    \big[ \sigma^x_0 \, U^{zz}_\otimes \big( (\epsilon-\textstyle \frac{J\Delta}{2})\frac\tau 2 \big) \big]^2 \,
    \big[ \sigma^z_0U^{xx}_\otimes \big( -\textstyle \frac{J(\gamma+1)\tau}{4} \big) \big]^2 \\
    & &
    \big[ \sigma^z_0 \, U^{yy}_\otimes \big( \textstyle \frac{J(\gamma-1)\tau}{4} \big) \big]^2\,
    U^{zz}_\otimes ( -\epsilon\tau ) \, U^z_0 (\omega_e \tau) \, U^z_\otimes \big( \textstyle \frac{J\lambda\tau}{2} \big)\nonumber
\earr
involving only global lattice displacements, global laser-induced
rotations and local addressing of atom $0$.

We note that an alternative scheme exists~\cite{duan03}, based on
tunnel coupling between neighboring atoms rather than on lattice
displacements, which can attain the simulation described here for
the special case $\gamma=0$ and would require some additional stroboscopic
steps in order to reproduce the general case.

Apart from 1D spin baths, this approach can be extended to other types
of environment. For example, it would be quite interesting
to consider, as an engineered bath, a 3D optical lattice.
Besides being feasible from an experimental point of view, this could be useful
in studying for instance the situation found in solid-state NMR~\cite{cory}.
It would be also intriguing to study the Bose-Hubbard model as a bath,
which would make the experimental realization even simpler~\cite{cucchietti06}.
Here we just focus on one-dimensional baths since, in several cases,
they are amenable to an exact solution.

\section{Results: the single-link scenario}
\label{sec:results}

In this Section we analyze the time evolution of 
the Loschmidt echo from Eq.~(\ref{eq:loschmidt}) for several distinct
scenarios where the qubit $S$ is coupled to
just one spin of the chain, i.e. $m=1$ in Eq.~\eqref{eq:hamint}.
In this case, in the thermodynamic limit the interaction
$\Ham_{\mathrm{int}}$ between the system $S$ and the environment $E$
does not affect the description of the bath Hamiltonian $\Ham_E$,
since it is local; therefore it can be considered in all senses
as a small perturbation of the environment.
The bath is effectively treated as a reservoir,
which is in contact with the system through just one point.
In the following we study the cases in which the bath is described
by a one dimensional spin-1/2 Ising, an $XY$ (Subsec.~\ref{subsec:XYanalyt})
and a Heisenberg (Subsec.~\ref{subsec:Heisenberg}) chain.

\subsection{$XY$-bath} 
\label{subsec:XYanalyt}

Here we focus on the case of a spin bath $E$ characterized by
a free Hamiltonian~(\ref{eq:spinbath}) of the $XY$ form,
i.e., with null anisotropy parameter along $z$ ($\Delta = 0$). 
In this case the environment is a one-dimensional spin-1/2 $XY$-model,
which is analytically solvable~\cite{lieb61}.
Below we show that also the Loschmidt echo can be evaluated
exactly~\cite{rossini06}, both for open and for periodic
boundary conditions.

In the case $m<N$ in which the system $S$ is coupled to just
some of the spins of the bath $E$, the perturbed Hamiltonian $\Ham_e$ of
Eq.~(\ref{eq:PERTHAM}) is the Hamiltonian of an $XY$ chain in a non-uniform
magnetic field. In this circumstance one cannot employ the approach
of Ref.~\cite{quan}, and in general the dynamical evolution of the system
has to be solved numerically.
The derivation we present here instead is analytical and it applies for all
values of $m=1,\cdots, N$. In the following we will present it for the case
of a generic $m$ but, in the remaining of the Section, we will explicitly
discuss its results only for the single-link case (i.e. $m=1$).

The first step of the analytical derivation is a Jordan-Wigner
transformation (JWT), in order to map both Hamiltonians $\Ham_g$
and $\Ham_e$ onto a free-Fermion model,
described by the quadratic form~\cite{lieb61}
\beq
     \Ham = \sum_{i,j} \left[ c_i^\dagger A_{i,j} c_j +
     \frac{1}{2} ( c_i^\dagger B_{i,j} c_j^\dagger + \mathrm{h.c.}) \right] +
     \frac{1}{2} \sum_i A_{i,i} \, ,
     \label{eq:quadratic}
\eeq
where $c_i, c_i^\dagger$ are the annihilation and creation operators
for the spinless Jordan-Wigner fermions, defined by
$$c_k = \exp \bigg( i \pi \sum_{j=1}^{k-1} \sigma_j^+ \sigma_j^- \bigg) \,
\sigma_k^-\;.$$
The two matrices {\bf A}, {\bf B} are given by
\barr
     \left[ {\bf A} \right]_{j,k} & = & - J \big( \delta_{k,j+1} +
     \delta_{j,k+1} \big) - 2 ( \lambda + \epsilon_j ) \delta_{j,k}, \\
     \left[ {\bf B} \right]_{j,k} & = & - \gamma J \big( \delta_{k,j+1} -
     \delta_{j,k+1} \big) \, ,
\earr
where $\epsilon_j = 0$ for $\Ham_g$, while
\beq
     \epsilon_j = \left\{ \begin{array}{ll} \epsilon &
     \: \textrm{if S is coupled to the } j \textrm{-th spin} \\
     0 & \: \textrm{elsewhere } 
     \end{array} \right.
\eeq
for $\Ham_e$.
A generic quadratic form, like Eq.~\eqref{eq:quadratic} (where {\bf A} is
a Hermitian matrix, due to the Hermiticity of $\Ham$,
and {\bf B} is antisymmetric, due to the anti-commutation rules among
the $c_i$), can be diagonalized in terms of the normal-mode spinless
Fermi operators $\{ \eta_k, \eta_k^\dagger \}$:
\beq
     \Ham = \sum_k E_k \left( \eta_k^\dagger \eta_k - \frac{1}{2} \right),
\eeq
where $\eta_k = \sum_i ( g_{k,i} c_i + h_{k,i} c_i^\dagger )$,
or in matrix form:
\beq
     \vec{\eta} = {\bf g} \cdot \vec{c} + {\bf h} \cdot \vec{c} \,\, {}^\dagger \, .
     \label{eq:eta_c}
\eeq
If we rewrite the two change-of-basis matrices ${\bf g}$ and ${\bf h}$
as $g_{k,i} \equiv \frac{1}{2} ( \phi_{k,i} + \psi_{k,i} )$
and $h_{k,i} \equiv \frac{1}{2} ( \phi_{k,i} - \psi_{k,i} )$,
we eventually arrive at the following coupled linear equations,
whose solution permits to find the eigenbasis of the non-uniform
Hamiltonian in Eq.~\eqref{eq:quadratic}:
\beq
     \left\{ \begin{array}{lcl}
     \vec{\phi}_k ( {\bf A} - {\bf B} ) & = & E_k \vec{\psi}_k, \\
     \vec{\psi}_k ( {\bf A} + {\bf B} ) & = & E_k \vec{\phi}_k.
     \end{array} \right. \label{eq:lin}
\eeq
Since ${\bf A}$ is symmetric and ${\bf B}$ is antisymmetric, all of the
$E_k$ are real; also the $g_{k,i}$ and the $h_{k,i}$ can be chosen
to be real.
The canonical commutation rules for the normal-mode operators impose the
constraints:
${\bf g} \, {\bf g^T} + {\bf h} \, {\bf h^T} = \openone$;
$\; {\bf g} \, {\bf h^T} - {\bf h} \, {\bf g^T} = 0$.

It is convenient to rewrite the spin-bath Hamiltonian-plus-interaction
$\Ham_e$ in the form
\beq
     \Ham_e = \frac{1}{2} {\bf \Psi^\dagger \, C \, \Psi}
     \label{eq:hame_c}
\eeq
where ${\bf \Psi^\dagger} = \left( c_1^\dagger \ldots c_N^\dagger \,
c_1 \ldots c_N \right)$ ($c_i$ are the corresponding spinless Jordan-Wigner
Fermion operators)
and ${\bf C} = \sigma^z \otimes {\bf A} + i \sigma^y \otimes {\bf B}$ is
a tridiagonal block matrix.

The Loschmidt echo, Eq.~\eqref{eq:loschmidt}, can then be evaluated
by means of the following formula~\cite{levitov96}:
\beq
     {\cal L}(t) = \left\vert \left< e^{-i t \sum_{i,j} C_{i,j} \Psi_i^\dagger \Psi_j}
     \right> \right\vert = \left\vert \mathrm{det}
     \left( 1 - {\bf r} + {\bf r} \, e^{-i {\bf C}t } \right) \right\vert \; ,
     \label{eq:det}
\eeq
where the elements of the matrix ${\bf r}$ are simply the two-point
correlation functions of the spin chain:
$r_{ij} = \langle G \vert \Psi^\dagger_i \Psi_j \vert G \rangle$,
where $\vert G \rangle$ is the ground state of $\Ham_g$.
An explicit expression of the correlators $r_{ij}$ as well as of the
matrix $e^{-i {\bf C} t}$ is given in Appendix~\ref{fermion}.

Equation~\eqref{eq:det} provides an explicit formula for the Loschmidt echo
in terms of the determinant of a $2N \times 2N$ matrix, whose entries
are completely determined by the diagonalization of the two linear systems
given by Eq.~\eqref{eq:lin}. This is one of the central results of our work.
It allows us to go beyond the central spin model where all the spins of $E$
are uniformly coupled with $S$ ($m=N$), whose solution (at least for
periodic boundary conditions) was discussed in~\cite{quan,cucchietti2,ou06}
and, for the sake of completeness, has been reviewed in Appendix~\ref{central}
(see Eq.~\eqref{eq:loschquan}). Notice that, similarly to the central
spin model, this formula allows to 
study a system composed of 
a large number of spins in the bath $N \sim 10^2 - 10^3$,
as it only requires manipulations of matrices whose size scales
linearly, and not exponentially, with $N$.

\subsubsection{Ising bath: general features}

\begin{widetext}

\begin{figure}[!h]
  \begin{center}
    \includegraphics[scale=0.55]{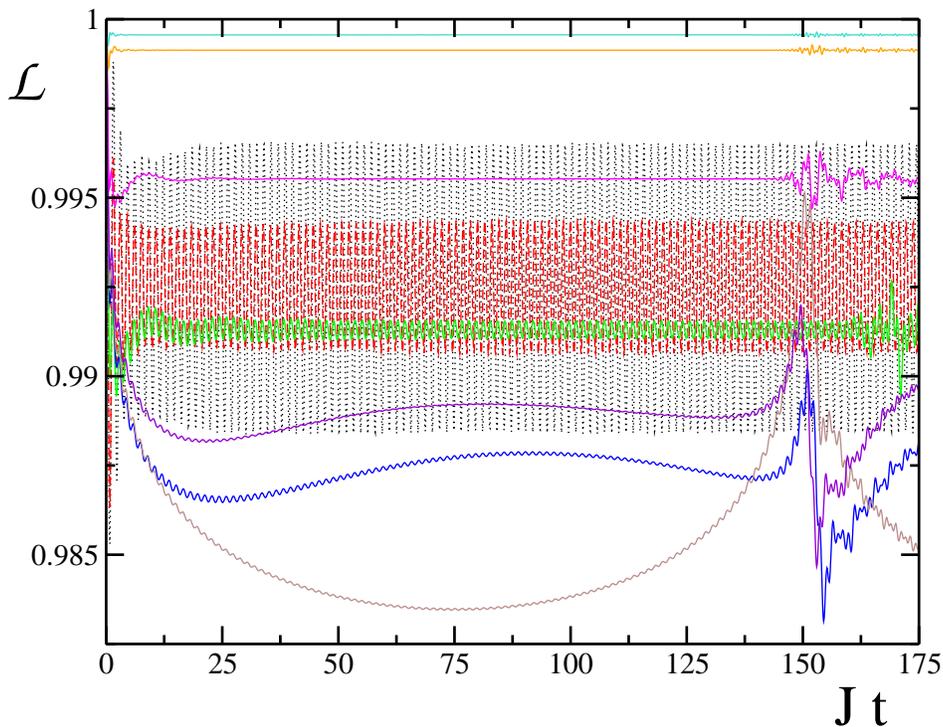}
    \caption{Single-link Ising model ($m=1,\Delta =0, \gamma=1$): 
      Loschmidt echo of Eq.~(\ref{eq:det}) as a function of time
      for a qubit coupled to a $N=300$ spin chain with periodic boundary
      conditions (here $\epsilon=0.25$).
      The various curves are for different values of the transverse
      magnetic field: $\lambda=$ 0.25 (black dotted), 0.5 (red), 0.9 (green),
      0.99 (blue), 1 (brown), 1.01 (violet), 1.1 (magenta), 1.5 (orange),
      1.75 (cyan). The critical point corresponds to $\lambda=1$
      (brown curve). Notice the revivals of quantum coherence
      at $t^* \sim 150$, due to the finite size of the chain~\cite{revivals}.}
    \label{fig:Ising1}
  \end{center}
\end{figure}

\end{widetext}

The generic behavior of $\mathcal{L}$ as a function of time for different 
values of $\lambda$, and fixed coupling constant $\epsilon$, is shown in
Fig.~\ref{fig:Ising1}.
For $\lambda < 1$ the echo oscillates with a frequency proportional
to $\epsilon$, while for $\lambda > 1$ the oscillation amplitudes
are drastically reduced. The Loschmidt echo reaches its minimum
value at the critical point $\lambda_c = 1$, thus revealing that
the decoherence is enhanced by the criticality of the environment.
Since the chain is finite, at long times there are revivals
of coherence~\cite{revivals}; in the thermodynamic limit $N \to \infty$
they completely disappear.
In any case, as it can be seen from the figure, already for $N=300$
spins there is a wide interval where the asymptotic behavior at long times
can be analyzed.

\subsubsection{Short-time behavior}

\begin{figure}[!h]
  \begin{center}
    \includegraphics[scale=0.315]{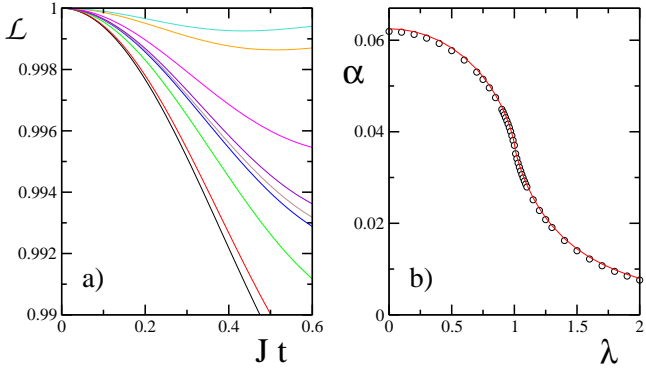}
    \caption{Short-time behavior of the Loschmidt echo:
      a) magnification of the plot in Fig.~\ref{fig:Ising1} for $t \leq 0.6$,
      where a Gaussian decay $L \sim e^{-\alpha t^2}$ is visible;
      b) dependence of $\alpha$ on $\lambda$ (the solid red line
      shows the perturbative estimate given by Eq.~\eqref{eq:timepert}).}
    \label{fig:Ising2}
  \end{center}
\end{figure}

At short times the Loschmidt echo $\mathcal{L}$ decays as a
Gaussian \cite{peres84}:
\beq
     \mathcal{L} (t) \sim e^{-\alpha t^2} \, ,
     \label{eq:smalldecay}
\eeq
as it can be seen in Fig.~\ref{fig:Ising2}a, which shows a magnification
of the curves from Fig.~\ref{fig:Ising1} for small times.
This behavior can be predicted within a second-order time
perturbation theory in the coupling $\epsilon$ between the system
and the bath: if  $\epsilon$ is small as compared to
the interaction $J$ between neighboring spins in the bath ($\epsilon \ll J$),
the decoherence factor $D(t)$ of Eq.~\eqref{eq:loschdef}  can be expanded
in series of $\epsilon$:
\barr
     \nonumber \langle e^{i \Ham_g t} \, e^{-i \Ham_e t} \rangle &
     = & \left\langle \mathcal{T} \left[ \exp \left(-i \int_0^t dt' e^{i \Ham_g t'}
     \mathcal{W} e^{-i \Ham_g t'} \right) \right] \right\rangle \\
     {} & \simeq & 1 + \epsilon \lambda_1 + \epsilon^\mathrm{2} \lambda_2 \, ,
     \label{eq:timeperturb}
\earr
where $\mathcal{T}$ is the time ordered product and
$\mathcal{W} = \epsilon \, \sigma^z_1$ accounts for the interaction
of the two-level system with the spin chain.
The above expression has to be evaluated on the ground state of $\Ham_g$,
therefore it is useful to rewrite the interaction $\mathcal{W}$ in terms
of the normal mode operators $\eta_k ^{(g)}$ of $\Ham_g$:
$$
  \begin{array}{rl}
  \mathcal{W} = &  \displaystyle \epsilon \bigg[ 2 \sum_{i,j}
    \big( g_{i,1} \eta_i^{(g)} {}^\dagger + h_{i,1} \eta_i^{(g)} \big) \cdot \\
    {} & \hspace{1.05cm} \cdot \displaystyle \big( g_{j,1} \eta_j^{(g)} +
    h_{j,1} \eta_j^{(g)} {}^\dagger \big) - 1 \bigg] \; .
  \end{array}
$$
The first-order term then reads
\beq
     \lambda_1 = -i t \bigg( 2 \sum_i \vert h_{i,1} \vert^2 - 1 \bigg) \, ,
\eeq
while the second-order term is given by
\barr
     \nonumber \lambda_2 & = & - \int_0^t d t'
     \int_0^{t'} d t'' \bigg[ 4 \sum_{i \neq j} \Big( (g_{i,1} h_{j,1})^2 \\
     \nonumber & - & g_{i,1} g_{j,1} h_{i,1} h_{j,1} \Big) \,
     e^{-i (E_i + E_j) (t' - t'')} \\
     & + & \Big( 2 \sum_i \vert h_{i,1} \vert^2 -1 \Big)^2 \bigg].
\earr
The Loschmidt echo is then evaluated by taking the square modulus
of the decoherence factor:
\beq
     \mathcal{L} (t) \simeq 1 - 4 \epsilon^2 t^2
     \sum_{i \neq j} \Big[ (g_{i,1} h_{j,1})^2 - g_{i,1} g_{j,1} h_{i,1} h_{j,1} \Big] \, .
     \label{eq:timepert}
\eeq

\begin{figure}[!h]
  \begin{center}
    \includegraphics[scale=0.32]{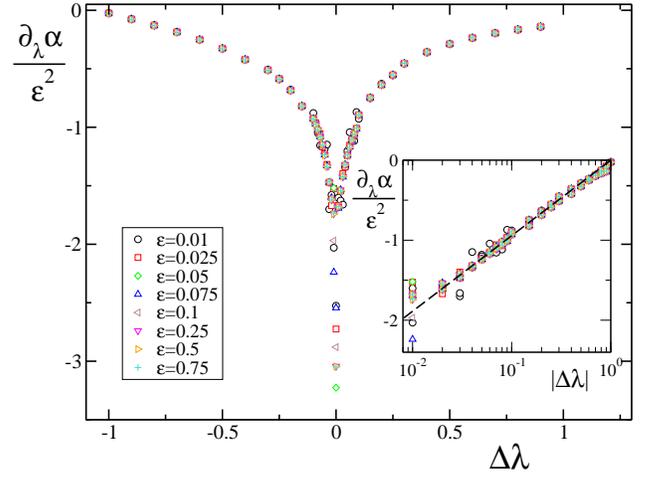}
    \caption{Single-link Ising model ($m=1,\Delta =0, \gamma=1$):
      Behavior of the Loschmidt echo of Eq.~(\ref{eq:det}) at short times.
      The plot shows the rescaled parameter
      $\partial_\lambda \alpha / \epsilon^2$
      as a function of $\lambda$ for a periodic $N=200$ Ising spin chain.
      Various symbols are for different values
      of the coupling strength $\epsilon$.
      Inset: plot in a semi-logarithmic scale.
      The dashed line indicates a fit of numerical data:
      $\partial_\lambda \alpha / \epsilon^2 =
      \tilde{c}_1 \log \vert \Delta \lambda \vert + \tilde{c}_2$,
      with $\tilde{c}_1 \approx 0.40983$ and $\tilde{c}_2 \approx 0.000108$.}
    \label{fig:Ising_Alpha}
  \end{center}
\end{figure}

In Fig.~\ref{fig:Ising2}b the initial Gaussian rate $\alpha$
is plotted as a function of $\lambda$; circles represent numerical data,
while the solid red curve is the perturbative estimate obtained from
second-order perturbation theory, given by Eq.~\eqref{eq:timepert}.
In Fig.~\ref{fig:Ising_Alpha} we analyze the behavior of
the first derivative of the rate $\alpha (\lambda, \epsilon)$
as a function of the distance from criticality
$\Delta \lambda \equiv \lambda - \lambda_c$, for a fixed
number $N$ of spins in the chain.
As predicted by the perturbative estimate, $\alpha$ scales like $\epsilon^2$;
most remarkably, its first derivative with respect to the transverse
field diverges if the environment is at the critical point $\lambda_c$.
In the inset we show that $\partial_\lambda \alpha$ diverges logarithmically
on approaching the critical value, as:
\beq
     \frac{\partial \alpha}{\partial \lambda} =
     c_1 \ln \vert \lambda - \lambda_c \vert + \rm{const.}
\eeq
This is a universal feature, entirely due to the underlying criticality
of the Ising model.

Our results show that at short times the Loschmidt echo is regular even in 
the presence of a bath undergoing a phase transition. The critical properties 
of the bath manifest in the {\em changes} of ${\cal L}$ when the bath 
approaches the critical point.

\subsubsection{Long-time behavior}

\begin{figure}[!h]
  \begin{center}
    \includegraphics[scale=0.3]{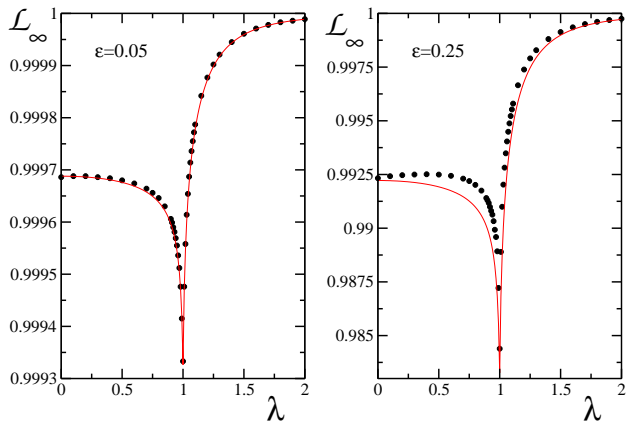}
    \caption{Single-link Ising model ($m=1,\Delta =0, \gamma=1$):
      Long-time behavior of the Loschmidt echo.
      The two plots show the saturation value $\mathcal{L}_\infty$
      as a function of $\lambda$ in an $N=200$ Ising spin chain
      with periodic b.c., for two different values
      of the coupling strength: $\epsilon=0.05 , \: 0.25$.
      Circles indicate numerical data, solid line is obtained from
      the perturbative calculation in Eq.~\eqref{eq:loschperturb}. }
    \label{fig:Ising_Plateau}
  \end{center}
\end{figure}

At long times, for $\lambda > 1 $ the Loschmidt echo approaches
an asymptotic value $\mathcal{L}_\infty$,
while for $\lambda < 1$ it oscillates around a constant value
(see Fig.~\ref{fig:Ising1} for a qualitative picture).
This limit value $\mathcal{L}_\infty$ strongly depends on $\lambda$
and presents a cusp at the critical point,
as shown in Fig.~\ref{fig:Ising_Plateau}.
%
%
Evidence that $\mathcal{L}_\infty$ describes the asymptotic regime
can be obtained by comparing data with the result of an analytical expression
based on the following simple ansatz:
We assume that at the thermodynamic limit $N \to \infty$ the coherence loss
saturates and is constant after a given transient time $t_0$
\beq
     \mathcal{L} \approx \mathcal{L}_\infty \qquad \forall \: t \geq t_0 .
     \label{eq:ansatz}
\eeq 
By writing the ground state $\ket{G}$ of $\Ham_g$ on the eigenbasis
of $\Ham_e$, the Loschmidt echo in Eq.~\eqref{eq:loschmidt} can be
formally written as
\beq
\mathcal{L} =
\Big\vert \sum_k \vert c_k \vert^2 e^{-i \, E_k^{(e)} t} \Big\vert^2 \;.
\eeq
Here, $c_k$ are the coefficients of the expansion
$\vert \psi_0^{(g)} \rangle =
\sum_k c_k \, \vert \psi_k^{(e)} \rangle$, where
$\Ham_e \vert \psi_k^{(e)} \rangle = E_k^{(e)} \vert \psi_k^{(e)} \rangle$
and $\vert \psi_0^{(g)} \rangle \equiv \ket{G}$.
An integration of the ansatz Eq.~(\ref{eq:ansatz})
from time $T > t_0$ to time $T+\Delta$ thus gives:
\beq
     \mathcal{L}_\infty  = \left| \sum_k \vert c_k \vert^2 e^{-i E_k^{(e)}
     \left( T + \frac{\Delta}{2} \right)} \:
     \textrm{sinc} \Big( \frac{E_k^{(e)} \Delta}{2} \Big) \right|^2 \, ,
\eeq
where $\textrm{sinc} \, (x) = \sin (x)/x$.
If the ground state in the perturbed Hamiltonian $\Ham_e$ is not degenerate,
we can just retain the term in the sum corresponding to $k=0$, thus obtaining
$\mathcal{L}_\infty \approx \vert c_0 \vert^4$.
Therefore a rough estimate of the Loschmidt echo is given by the overlap
between the ground state of $\Ham_g$ and the ground state of $\Ham_e$
(see Ref.~\cite{zanardi06} for a detailed analysis of the ground
state fidelity in spin systems).
This can be evaluated via second-order perturbation theory
in $\epsilon$:
\beq
     c_0 = 1 - \frac{1}{2} \sum_{k \neq 0} \frac{\left\vert \langle \psi_k^{(g)}
     \vert \, \mathcal{W} \, \vert \psi_0^{(g)} \rangle \right\vert^2}
     {\left( E_k^{(g)} - E_0^{(g)} \right)^2 } \label{eq:perturb}
\eeq
where $\vert \psi_k^{(g)} \rangle$ are the excited states of $\Ham_g$
with energies $E_k^{(g)}$ and
$\mathcal{W} = \epsilon \sum_{j=j_1}^{j_m} \sigma^z_1$.
The expression in Eq.~\eqref{eq:perturb} can be easily computed for
the case $m=1$ following the same steps used to evaluate
Eq.~\eqref{eq:timeperturb}. The final result is:
\beq
     \mathcal{L}_\infty \approx \Bigg[ 1 - 2 \epsilon^2 \sum_{i \neq j}
     \bigg( \frac{g_{i1}
     \, h_{j1}}{\tilde{E}_i^{(g)} + \tilde{E}_j^{(g)}} \bigg)^2 \, \Bigg]^4 ,
     \label{eq:loschperturb}
\eeq
where the energies $\tilde{E}_k^{(g)}$ are rescaled on the ground state
energy, $\tilde{E}_k^{(g)} = E_k^{(g)} - E_0^{(g)}$.
In Fig.~\ref{fig:Ising_Plateau} we plotted these results for two
different values of the perturbation strength (solid lines);
they can be compared with the exact numerical results (circles).
Notice that analytic estimates are more accurate for small $\epsilon$,
and far from the critical region.

\subsubsection{Ising bath: scaling with the size}

\begin{figure}[!h]
  \begin{center}
    \includegraphics[scale=0.7]{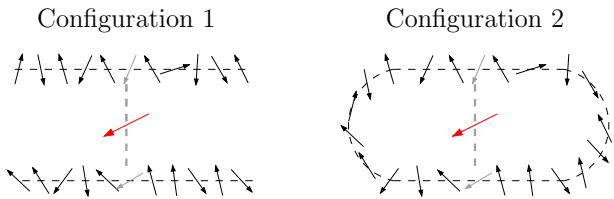}
    \caption{Qubit (red) coupled to two non-communicating $N$-sites
      chains (configuration 1) and to two spins of a $2N$-sites chain
      (configuration 2). Here, as an example, we set $N=11$.}
    \label{fig:chains}
  \end{center}
\end{figure}

As one could expect, contrary to the case of a non-local system-bath
coupling~\cite{quan}, if the environment is not at the critical point,
the decay of the coherences is independent of the number of spins in the bath.
This is due to the fact that correlations away from criticality decay
exponentially, while they decay as a power law at the critical point.
In particular, we checked that both the initial rate $\alpha$ and the
saturation value $\mathcal{L}_\infty$ do not depend significantly
on $N$ for $\lambda \neq \lambda_c$.
We present here a simple argument as a confirmation of this statement.
Far from the critical region, independently of the bath size
(for sufficiently large baths),
a qubit simultaneously coupled to two non-communicating $N$-sites chains
via a one-spin link (configuration 1) should decohere in the same way
as if it was coupled to two sufficiently distant spins of a $2N$-sites
chain (configuration 2).
A pictorial representation of the two configurations described above
is shown in Fig.~\ref{fig:chains}, where the red spin represents the
system, the grey ones are the bath spins directly coupled to the system.

\begin{widetext}

\begin{figure}[!h]
  \begin{center}
    \includegraphics[scale=0.5]{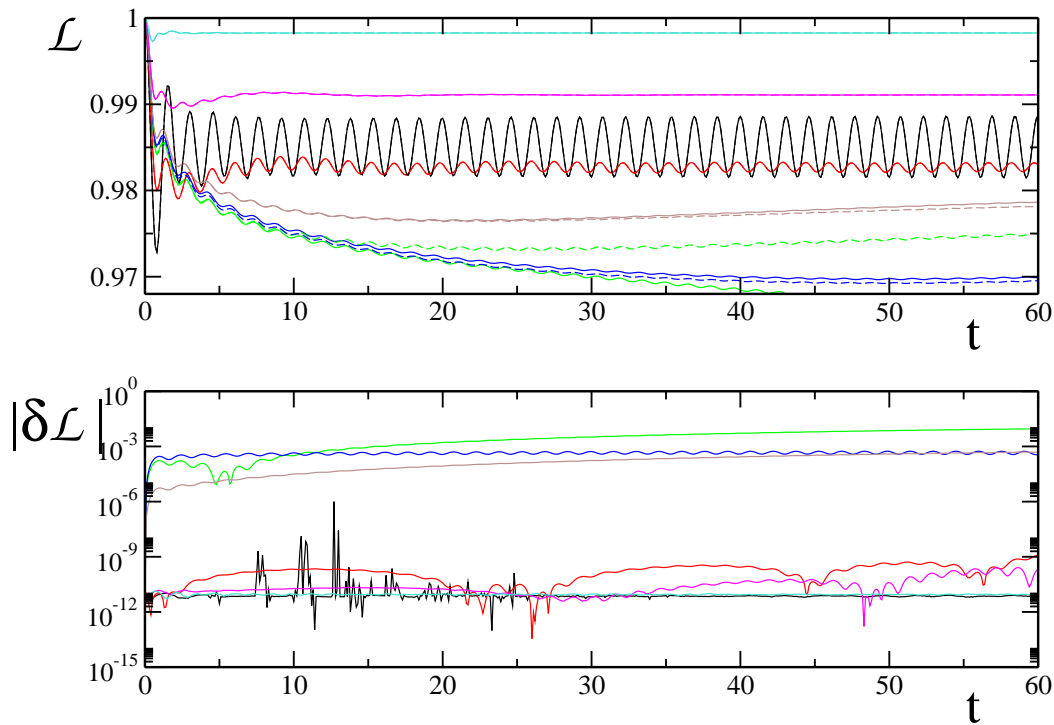}
    \caption{Upper figure: solid curves stand for the qubit coupled to
      two non communicating open $N=200$ spin Ising chains (configuration 1);
      dashed curves are for the qubit coupled to two opposite spins
      of a periodic $N=400$ spin Ising chain (configuration 2).
      Different curves are for various values of the transverse magnetic field:
      $\lambda=0.5$ (black), 0.9 (red), 0.99 (green), 1 (blue), 1.01 (brown),
      1.1 (magenta), 1.5 (cyan).
      Lower figure: absolute differences between the two configurations.}
    \label{fig:Ising_open_middle}
  \end{center}
\end{figure}

\end{widetext}

We numerically checked that, far from the critical region,
the two configurations leave the decoherence unchanged.
In particular we chose an $N=200$ chain, and obtained that,
if $\vert \lambda - \lambda_c \vert \gtrsim 10^{-2}$,
the Loschmidt echo evaluated in the two configurations is the same,
up to a discrepancy of $10^{-5}$, as it can be seen
from Fig.~\ref{fig:Ising_open_middle}.

\begin{figure}[!h]
  \begin{center}
    \includegraphics[scale=0.34]{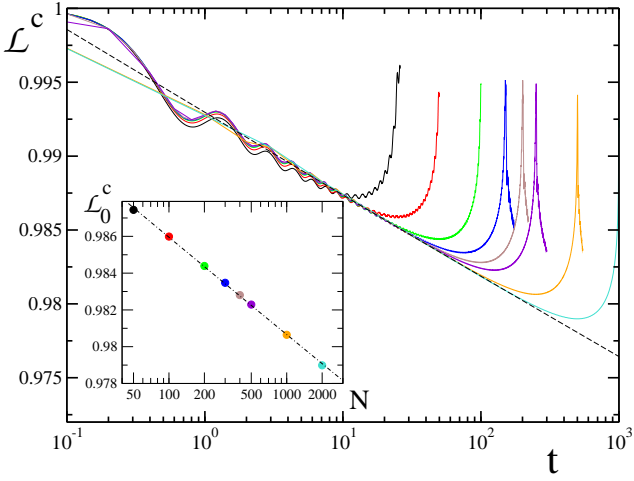}
    \caption{Single-link Ising model ($m=1,\Delta =0, \gamma=1$).
      Loschmidt echo as a function of time at the critical point
      $\lambda = 1$, for different sizes of the periodic chain: $N=50$ (black),
      100 (red), 200 (green), 300 (blue), 400 (brown), 500 (violet),
      1000 (orange), 2000 (cyan). The perturbation strength is kept fixed:
      $\epsilon = 0.25$. The dashed line is a guideline that shows a decay
      of the type $\mathcal{L}^c(t) = c_0 / (1+ c_1 \, \textrm{ln} \, t)$.
      Inset: Minimum value of $\mathcal{L}^c$ as a function of $N$.
      Numerical data have been fitted with
      $\mathcal{L}_0^c = \mathcal{L}_0 / (1 + \beta \, \textrm{ln} N)$
      (dot-dashed line), where
      $\mathcal{L}_0 \approx 0.99671,\; \beta \approx 2.36933 \times 10^{-3}$.}
    \label{fig:Ising_Critic}
  \end{center}
\end{figure}

At the critical point instead the behavior dramatically changes.
In Fig.~\ref{fig:Ising_Critic} we plot the decay of $\mathcal{L}$
in time at $\lambda_c$ for different sizes of the chain
and fixed perturbation strength, $\epsilon = 0.25$.
The minimum value $\mathcal{L}_0^c$ reached by the Loschmidt echo decays
with the size $N$ as
\beq
     \mathcal{L}_0^c = \frac{\mathcal{L}_0}{1+ \beta \ln (N)}.
     \label{eq:scaling}
\eeq
Notice also that revivals are due to the finiteness of the environmental
size (as in Fig.~\ref{fig:Ising1})~\cite{revivals}.
We expect that the coherence loss should go to zero at the thermodynamic
limit $N \to \infty$; this is hard to see numerically, since the decay
is logarithmic, and the actual value of $\mathcal{L}_0^c$ is still
quite far from zero, even for $N=2000$ spins.

\subsubsection{$XY$-baths with arbitrary $xy$-anisotropy $\gamma$} 
\label{subsec:XX}

\begin{figure}[!h]
  \begin{center}
    \includegraphics[scale=0.315]{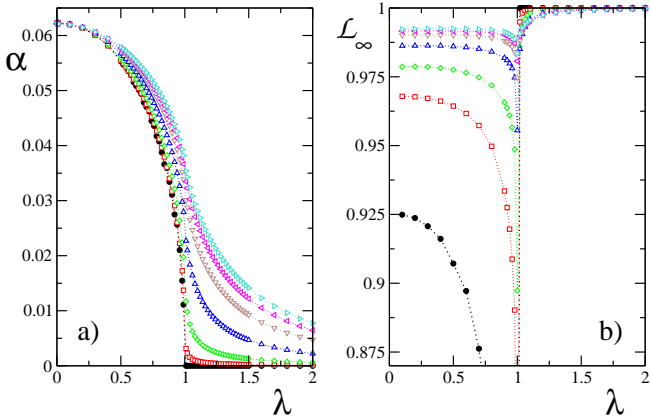}
    \caption{Single-link $XY$-models ($m=1,\Delta =0$):
      Short- and long-time behavior of the Loschmidt echo
      for a qubit coupled to a periodic $N=300$ spin $XY$ chain,
      with a coupling strength $\epsilon = 0.25$.
      a) Small times (initial Gaussian decay rate $\alpha$); b) Long times
      (saturation values ${\cal L}_\infty$).
      The various curves are for different anisotropy values:
      $\gamma=$ 0 ($XX$-model, full black circles), 0.1 (red squares),
      0.25 (green diamonds), 0.5 (blue triangles up),
      0.75 (brown triangles down), 0.9 (magenta triangles left),
      1 (Ising, cyan triangles right).}
    \label{fig:XY_Alpha_Plateau}
  \end{center}
\end{figure}

The properties described in the previous Subsections are typical
of the Ising universality class, indeed they remain qualitatively the same
as far as $\Delta = 0, \gamma \neq 0$ in Eq.~\eqref{eq:spinbath}: 
in particular this class of models has one critical point at $\lambda =1$.
Following our previous analysis of the Ising model ($\gamma=1$),
we focus on the two distinct regions of short- and long-time behavior,
whose features are depicted in Fig.~\ref{fig:XY_Alpha_Plateau}.

At small times the decay is again Gaussian, as in Eq.~\eqref{eq:smalldecay},
and the first derivative of the decay rate $\alpha$ with respect to $\lambda$
diverges at $\lambda_c$ (see Fig~\ref{fig:XY_Alpha_Plateau}a).
As far as the $xy$ anisotropy decreases, the system approaches a limiting
case in which $\partial_\lambda^- \alpha = -\infty$, while
$\partial_\lambda^+ \alpha = 0$.
The asymptotic value $\mathcal{L}_\infty$ at long times shows a cusp
at $\lambda = 1$; notice that, as the anisotropy $\gamma$ decreases,
despite the fact that the decay of coherences at short times is always reduced,
their saturation value $\mathcal{L}_\infty$ becomes lower for $\lambda <1$,
while gets higher for $\lambda >1$. Therefore decoherence is asymptotically
suppressed only in the ferromagnetic phase, while it is enhanced
for $\lambda <1$.
As we said before, when $\gamma \to 0$ the system is driven towards
a discontinuity at $\lambda=1$. This is due to the fact that for $\lambda > 1$
the ground state approaches a fully polarized ferromagnetic state.
Finally we analyzed the scaling of the decoherence with the size
of the environment at criticality.
As in the Ising model, the minimum value $\mathcal{L}_0^c$
reached by the Loschmidt echo at $\lambda_c$ depends on the
chain length $N$ as in Eq.~\eqref{eq:scaling}:
$\mathcal{L}_0^c = \mathcal{L}_0/(1+ \beta \ln (N))$.

\begin{figure}[!h]
  \begin{center}
    \includegraphics[scale=0.3]{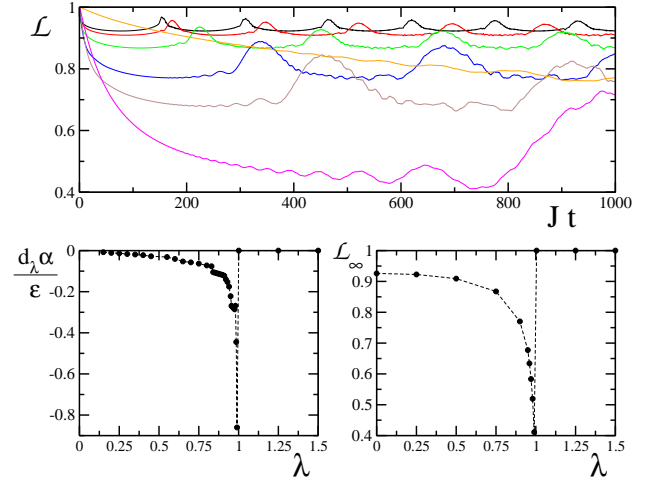}
    \caption{Single-link $XX$-model ($m=1,\Delta = \gamma = 0$):
      Temporal behavior of the Loschmidt echo for a qubit coupled
      to a periodic $N=300$ spin isotropic $XX$ chain; $\epsilon = 0.25$.
      The various curves stand for different values of the magnetic field:
      $\lambda=$ 0.25 (black), 0.5 (red), 0.75 (green), 0.9 (blue),
      0.95 (brown), 0.99 (magenta), 1. (orange).
      Bottom left: first derivative of the decay rate $\alpha$ at short times
      with respect to $\lambda$, as a function of $\lambda$.
      Bottom right: saturation value $\mathcal{L}_\infty$ at long times.}
    \label{fig:XXmodel}
  \end{center}
\end{figure}

The system behaviour dramatically changes when it becomes isotropic, i.e. at
$\gamma = 0$. In this case the environment is an $XX$ spin chain and
it does not belong any more to the Ising universality class.
In particular it exhibits a critical behavior over the whole parameter range
$\vert \lambda \vert \leq 1$, while it is ferromagnetic (anti-ferromagnetic)
for $\lambda > 1$ ($\lambda < -1$).
The decay of the Loschmidt echo (shown in Fig.~\ref{fig:XXmodel}) reflects
its critical properties:
indeed we found that it behaves as in Eq.~\eqref{eq:scaling}
over the whole range $\vert \lambda \vert \leq 1$.
In the ferromagnetic case instead the coupled qubit does not decohere
at all (i.e. $\mathcal{L} (t) = 1$), since the ground state
of the $XX$-model for $\lambda > 1$ is the fully polarized state
with all spins parallel to the external field.
The first derivative of the initial decay rate $\partial_\lambda \alpha$
is monotonically negative in the critical region by increasing $\lambda$ and
diverges for $\lambda \to 1^{-}$, while it is strictly zero for $\lambda > 1$.
Also the plateau $\mathcal{L}_\infty$ as a function of $\lambda$
presents a discontinuity in $\lambda = 1$, since it drops to zero
for $\lambda \to 1^{-}$, and it equals one for $\lambda > 1$.

\subsection{$XXZ$-Heisenberg chain bath} 
\label{subsec:Heisenberg}

In this Section we consider  a single-link ($m=1$) $XXZ$ anisotropic Heisenberg
chain ($\lambda = \gamma = 0$ in Eq.~\eqref{eq:spinbath})
with anisotropy parameter $\Delta$.
We resort to the numerical t-DMRG  to compute the Loschmidt
echo~(\ref{eq:loschmidt}), since this specific
spin model is not integrable. A brief introduction to the static and 
the time-dependent DMRG algorithm is given in Appendix~\ref{DMRG}.
We consider open boundary conditions, since DMRG with periodic boundaries
intrinsically gives much less accurate results~\cite{schollwock}.
The evaluation of the Loschmidt echo with the t-DRMG is more time- and 
memory-consuming than its exact calculation in the solvable cases, therefore
at present we could not study the behavior of $\mathcal{L} (t)$ at long times.
Nonetheless we are able to fully analyze the short-time behavior in
systems with environment size of up to $N \sim 10^2$ spins.
In the following we show the results concerning the coupling of the two-level
system to one spin in the middle of the chain: this corresponds to the case
with less border effects. We numerically checked that the results are
not  qualitatively affected from this choice, but changing the
system-environment link position results in a faster appearance 
of finite size effects due to the open boundaries.

\begin{figure}[!h]
  \begin{center}
    \includegraphics[scale=0.33]{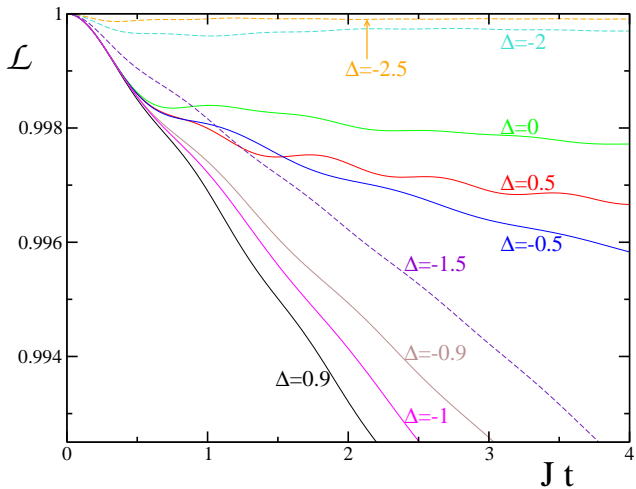}
    \caption{Single-link $XXZ$-model ($m=1,\lambda = \gamma = 0$):
      Loschmidt echo as a function of time for a qubit coupled
      to the central spin of an open ended $N=100$ spin $XXZ$-Heisenberg
      chain, with coupling strength $\epsilon=0.1$.
      The various curves are for different values of the anisotropy:
      $\Delta=0.9$ (black), 0.5 (red), 0 (green), -0.5 (blue), -0.9 (brown),
      -1 (magenta), -1.5 (violet), -2 (cyan), -2.5 (orange); curves
      corresponding to non-critical situations are dashed.}
    \label{fig:Heis}
  \end{center}
\end{figure}

\begin{figure}[!h]
  \begin{center}
    \includegraphics[scale=0.37]{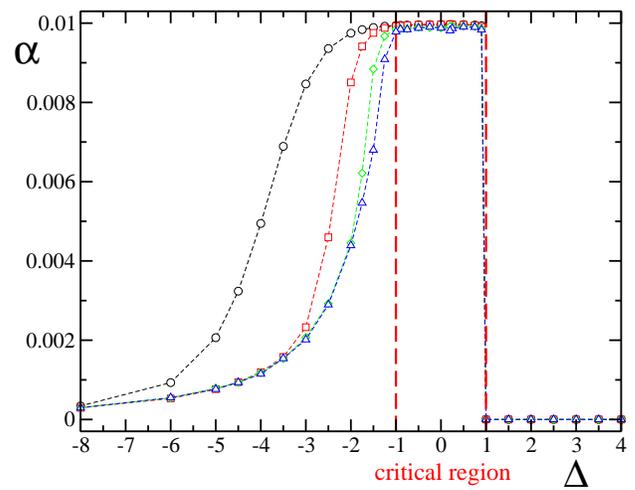}
    \caption{Scaling of the decay rate $\alpha$ at short times
      for an open ended $XXZ$-Heisenberg chain as a function of $\Delta$.
      The various symbols are for different sizes of the chain:
      $N=10$ (black circles), 20 (red squares), 50 (green diamonds),
      100 (blue triangles). The coupling strength between the qubit
      and the chain is kept fixed at $\epsilon = 0.1$.}
    \label{fig:Heis_Alpha}
  \end{center}
\end{figure}

In Fig.~\ref{fig:Heis} we plot the decay of the Loschmidt
echo as a function of time, for various values of the anisotropy $\Delta$
and fixed coupling strength $\epsilon = 0.1$. 
In the ferromagnetic zone outside of the critical region
($\Delta \ge 1$) $\mathcal{L}$ does not decay at all, as a consequence
of the ground state full polarization. 
In the critical region and for long times, the Loschmidt echo decay is
proportional to the modulus of $\Delta$ and  for $\Delta > -1$ it slows down
until it is completely suppressed in the perfectly anti-ferromagnetic regime
($\Delta \to - \infty$).
The short time decay is again Gaussian at short times and the rate $\alpha$
is shown in Fig.~\ref{fig:Heis_Alpha} for various sizes of the bath.
We notice two qualitatively different behaviors at the boundaries
of the critical region: at $\Delta=+1$ there is a sharp discontinuity,
while at $\Delta=-1$ the curve is continuous.
In the critical region $-1 \leq \Delta \leq 1$ the initial decay rate $\alpha$
is constant and reaches its maximum value due to the presence of low energy
modes, while in the ferromagnetic phase $\Delta \ge 1$ it is strictly zero.
In the figure, finite size effects are evident: indeed, contrary to
the $XY$-model, the decay rate $\alpha$ changes with the bath size $N$
outside the critical region.
However, as the system approaches the thermodynamic limit ($N \to \infty$),
the dependence on $N$ weakens and, while at $\Delta=-1$ the curve
$\alpha (\Delta)$ appears to remain continuous, its first derivative
with respect to $\Delta$ tends to diverge. On the contrary, the
ferromagnetic transition $\Delta = 1$ is discontinuous independently of $N$.

\subsection{Decoherence and entanglement}
\label{sec:deco_ent}

We propose here to establish a link between decoherence effects on the system
and entanglement inside the environment.
This study can be justified by the fact that decoherence properties
of the qubit system seem to be intrinsically related to quantum
correlations of the bath, as the proximity to critical points reveals.
On the other hand, entanglement quantifies the amount of these
correlations that do not have classical counterparts,
and it has been widely studied in the recent years, especially
in connection with the onset of quantum phase
transitions~\cite{osterloh02,osborne02,vidal03,stauber04}.

In Ref.~\cite{dawson} it was shown that two-party entanglement
in the environmental bath of a central spin model can suppress
decoherence; this effect has been explained as a consequence of
entanglement sharing, and it was supposed to be common to any system
whose environment maintains appreciable internal entanglement, while
evolving in time.
We now characterize a more complex case of system-plus-bath coupling,
given by Eq.~\eqref{eq:system+bath}, with a 
richer structure in the ground state entanglement
which suggests the following picture, valid at short times for $J t \ll 1$.

We expect that when the decay of coherences
is quadratic, only short range correlations in the bath are
important, therefore it seems natural to relate the short-time decay rate
$\alpha$ to the two-site nearest neighbor entanglement of the chain.
All the information needed to analyze bipartite entanglement between
two spins inside the chain, say $i$ and $j$, is contained
in the two-qubit reduced density matrix $\rho_{ij}$, obtained after
tracing out all the other spins.
We use the concurrence $C(\vert i-j \vert)$ as an entanglement measure
for arbitrary mixed state $\rho_{ij}$ of two qubits  defined as \cite{wootters}
\beq
     C(\vert i-j \vert) = \max ( \lambda_1 - \lambda_2 - \lambda_3 - \lambda_4, 0 ),
\eeq
where $\lambda_1$ are the square roots of the eigenvalues of the matrix
$\rho_{ij} \tilde{\rho}_{ij}$, in decreasing order;
the spin flipped matrix of $\rho_{ij}$ is defined as
$\tilde{\rho}_{ij} = (\sigma_y \otimes \sigma_y) \, \rho^*_{ij} \,
(\sigma_y \otimes \sigma_y)$, and the complex conjugate is computed in
the canonical basis $\{ \vert \uparrow \uparrow \rangle,
\vert \uparrow \downarrow \rangle, \vert \downarrow \uparrow \rangle,
\vert \downarrow \downarrow \rangle \}$ .
The entanglement of formation of the mixed state $\rho_{ij}$ is 
a monotonic function of the concurrence.

We start our analysis by considering again the $XY$ spin bath: 
the concurrence $C(k)$ in terms of one-point and two-point spin-correlation
functions can be analytically evaluated~\cite{amico04,barouch}.
As long as $\gamma \ne 0$, this system belongs to the Ising universality
class, for which it has been shown that the concurrence between two spins
vanishes unless the two sites are at most
next-to-nearest neighbor~\cite{osterloh02,osborne02}.
The nearest-neighbor concurrence $C(1)$ presents a maximum
that occurs at $\lambda = \lambda_m > 1$, and it is not related
to the critical properties of the model.
For $\lambda > \lambda_m$ it monotonically decreases to zero,
as the ground state goes towards the ferromagnetic product state.
For $\lambda < \lambda_m$ the concurrence decreases and precisely reaches
zero at $\lambda_K = \sqrt{ (1+\gamma) (1-\gamma) }$,
where the ground state exactly factorizes~\cite{kurmann}.

\begin{figure}[!ht]
  \begin{center}
    \includegraphics[scale=0.33]{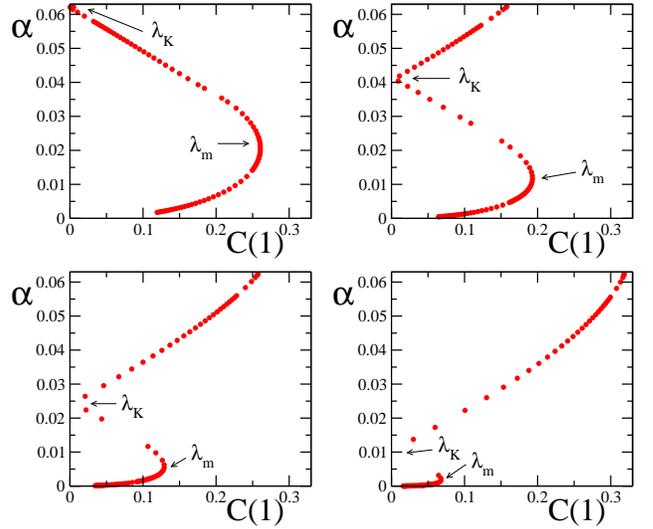}
    \caption{Dependence of the Loschmidt echo decay rate $\alpha$ on
      the nearest-neighbor concurrence $C(1)$ in the $XY$-model,
      for $\epsilon=0.25$ and different anisotropy values:
      $\gamma=1$ (upper left), 0.5 (upper right),
      0.25 (lower left), 0.1 (lower right).
      Data have been evaluated for an $N=301$ spin chain.}
    \label{fig:XY_Conc_Alpha}
  \end{center}
\end{figure}

In the proximity of the quantum phase transition, the entanglement
obeys a scaling behavior~\cite{osterloh02}; in particular at the
thermodynamic limit the critical point is characterized by a
logarithmic divergence of the first derivative of the concurrence:
$\partial_\lambda C (1) \sim \ln \vert \lambda - \lambda_c \vert$.
A similar divergence has been found for the first derivative of the
decay rate $\alpha$ of the Loschmidt echo (see the inset
of Fig.~\ref{fig:Ising_Alpha}).
In Fig.~\ref{fig:XY_Conc_Alpha} we analyzed the dependence
of $\alpha$ as a function of the nearest-neighbor concurrence $C(1)$.
The behavior $\alpha (C)$ is not monotonic,
since $\alpha$ monotonically decreases with $\lambda$
(see Fig.~\ref{fig:Ising2}b), while $C(1)$ has a non-monotonic behavior.
Indeed, for $\lambda > \lambda_m$ and $\lambda < \lambda_K$ the two 
quantities are correlated, while they are anti-correlated for 
$\lambda_m > \lambda > \lambda_K$.

\begin{figure}[!h]
  \begin{center}
    \includegraphics[scale=0.33]{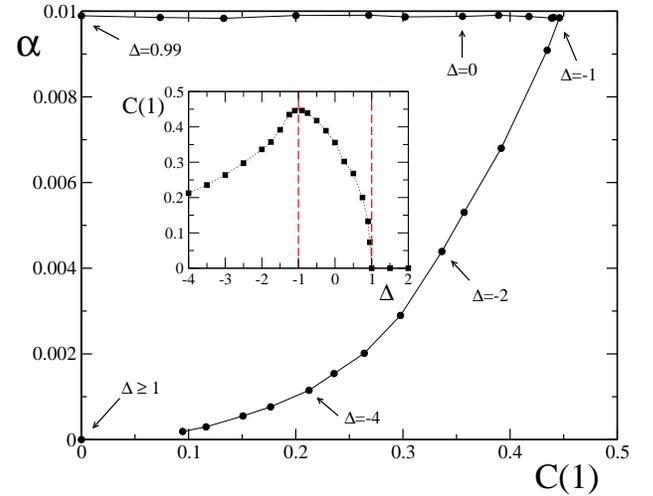}
    \caption{Decay rate $\alpha$ as a function of the nearest-neighbor
      concurrence $C(1)$ in the Heisenberg model for $N=100$, $\epsilon=0.1$.
      Inset: nearest-neighbor concurrence as a function of the anisotropy
      $\Delta$. The two dashed lines isolate the critical region.}
    \label{fig:Heis_Conc_Alpha}
  \end{center}
\end{figure}

A similar scenario has been found in the case of the Heisenberg chain
as a spin bath, reported in Fig.~\ref{fig:Heis_Conc_Alpha}. 
As for the Ising model, the behavior of $\alpha (C)$ is not monotonic,
following the concurrence behavior:
Indeed $C(1)$ is zero for $\Delta \geq 1$ (in the fully polarized
ground state), it monotonically increases while decreasing $\Delta$ up to
$\Delta=-1$, and then starts to decrease again, vanishing in the
perfect anti-ferromagnetic limit $\Delta \to -\infty$~\cite{syljuasen}.
In conclusion, the behavior $\alpha = \alpha(C)$ is monotonically increasing
in the anti-ferromagnetic region, is constant at the criticality,
and is strictly zero for the ferromagnetic phase
(see Figs.~\ref{fig:Heis}b,~\ref{fig:Heis_Alpha}),
resulting in a direct correlation of $\alpha$ and $C(1)$ in
the anti-ferromagnetic phase.

\section{Results: the multiple-links scenario}
\label{sec:manylinks}

\begin{figure}[!ht]
  \begin{center}
    \includegraphics[scale=0.8]{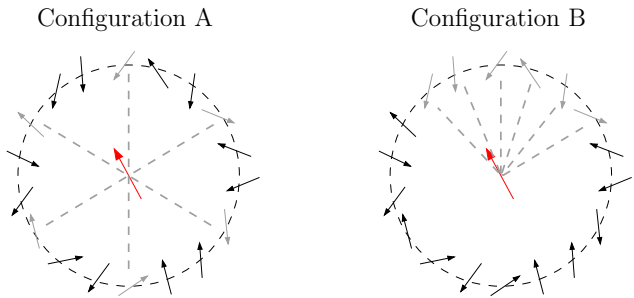}
    \caption{Different setups for the coupling of the two-level
      system (red spin) with the spin chain.
      The qubit is linked to the spins of the chain depicted in grey.
      Configuration A shows a star-symmetric setup, while configuration B
      is a non-symmetric packed setup.
      As a pictorial example, here we set $N=18, \, m=6$.}
    \label{fig:mvar_geometry}
  \end{center}
\end{figure}

In this Section we study the case in which
the system $S$ is non-locally coupled to more than one spin of the bath $E$,
namely we choose $m \neq 1$ in the interaction Hamiltonian
Eq.~\eqref{eq:hamint}.
In general, given a number $m$ of links, there are many
different ways to couple the two-level system with the bath spins.
In the following we will consider two geometrically different setups:
a spin-symmetric configuration (type A)
where the qubit is linked to some spins of the chain
which are equally spaced, and a non-symmetric configuration (type B)
where the spins to which the qubit is linked are nearest neighbor.
We will also suppose that the interaction strength $\epsilon$ between
the qubit and each spin of the bath is kept fixed and equal for all the links.
A schematic picture of the two configurations is given
in Fig.~\ref{fig:mvar_geometry}.
We will present results concerning an environment constituted by an Ising
spin chain with periodic boundary conditions.
Notice that this system-environment coupling can induce the emergence
of criticality in $E$ even for small couping $\epsilon \ll J$ in the 
central spin scheme, i.e. $m = N$~\cite{quan,cucchietti2,ou06}.
Indeed, in this case, a perturbation $\epsilon$ results in
a change of the external magnetic field of the chain from $\lambda$ to
$\lambda + \epsilon$.
If the environment is characterized by $\lambda <1$, at
$\epsilon > 1 - \lambda$ the coupling drives a quantum phase transition
in the bath~\cite{cucchietti2}.

\subsubsection{Short-time behavior}

\begin{figure}[!ht]
  \begin{center}
    \includegraphics[scale=0.31]{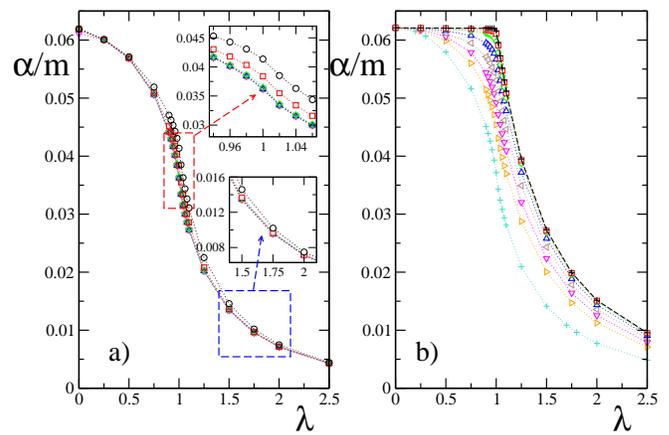}
    \caption{Decay rate $\alpha$ rescaled with respect to
      the number of links $m$, in a star symmetric configuration (type A)
      on the left, and in a non-symmetric configuration (type B) on
      the right.
      The environment is an $N=300$ spin Ising chain with periodic
      boundary conditions. The interaction strength $\epsilon$ is kept fixed
      for all the links between the qubit and the environment
      spins: $\epsilon=0.25$.
      Various symbols stand for different values of $m$:
      1 (plus), 2 (triangles right), 3 (triangles down), 5 (triangles left),
      10 (triangles up), 30 (diamonds), 100 (squares), 150 (circles),
      300 (dashed line, right figure).
      The inset on the left figure shows a magnification of the same plot,
      centered in proximity of the critical point $\lambda_c = 1$.}
    \label{fig:Ising_mvar_alpha}
  \end{center}
\end{figure}

At short times the Loschmidt echo exhibits a Gaussian decay,
as in Eq.~\eqref{eq:smalldecay}, for both configurations.
For the setup A the decay rate $\alpha$, far from the critical region,
scales as
\beq
     \alpha \propto m \, \epsilon^2,
     \label{scalconfA}
\eeq
as it can be seen from Fig.~\ref{fig:Ising_mvar_alpha}a.
The scaling with the number of links $m$ is a consequence of the fact that
the short-time behavior is dominated by the dynamics of the environment
spins close to those linked to the qubit (see also Sec.~\ref{sec:deco_ent}).
Therefore, if the linked spins are not close to each other,
they do not interact among them on the short time scale $J t \ll 1$.
Near the critical point this picture is not valid, since 
long-range correlations between the spins of the bath
become important, even at small times.
In this region indeed the scaling $\alpha \sim m$ is less appropriate,
as it can be seen in the inset of Fig~\ref{fig:Ising_mvar_alpha}a.

The scaling of Eq.(\ref{scalconfA}) does not hold any more in the
setup B, as in this case collective modes influence the dynamics
of the system even at short times, since the spins linked to the central
spin are close to each other and they are not independent any more.
This is clearly visible in Fig.~\ref{fig:Ising_mvar_alpha}b.
We notice also that, as $m/N$ increases, $\alpha$ tends to remain constant
for $\lambda \leq 1$ and then decreases for $\lambda >1$.
In the limiting case $m = N$, $\alpha$ is positive and strictly constant
for $\lambda \leq \lambda_c$; the first derivative $\partial_\lambda \alpha$
presents a discontinuity at $\lambda_c$, showing a divergence
from the ferromagnetic zone $\lambda \to 1^+$.

\subsubsection{Long-time behavior}

\begin{figure}[!ht]
  \begin{center}
    \includegraphics[scale=0.33]{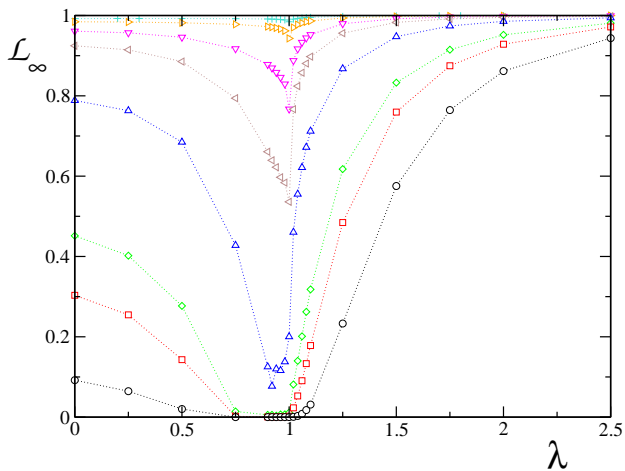}
    \caption{Asymptotic value of the Loschmidt echo for the setup B;
      the environment is a 300 spin Ising chain, with $\epsilon=0.25$.
      Various symbols stand for different values of $m$:
      1 (plus), 2 (triangles right), 5 (triangles down), 10 (triangles left),
      30 (triangles up), 100 (diamonds), 150 (squares), 300 (circles).}
    \label{fig:Ising_mvar_pack_long}
  \end{center}
\end{figure}

In this Section we concentrate on the setup B, since the configuration A
is less interesting: indeed for long times the Loschmidt echo behaves
as if the qubit was coupled to an environment with a smaller number of spins.
In Fig.~\ref{fig:Ising_mvar_pack_long} we plotted the asymptotic value
of the Loschmidt echo $\mathcal{L}_\infty$ for the setup B.
We observe that, as $m$ increases, the coherence loss enhances,
and the valley around the critical point $\lambda_c$ deepens and gets broader.
Notice also that, when approaching the central spin limit, the asymptotic value
$\mathcal{L}_\infty$ reaches values very close to zero, even far
from criticality. This situation is completely different from what occurs
in the single-link scenario where, away from criticality, the Loschmidt echo
remains very close to one even for non negligible system-bath coupling
strengths (e.g., for $\epsilon = J/4$ we found
$\mathcal{L}_\infty \sim 0.99$, as it can be seen, for example,
from Figs.~\ref{fig:Ising1},~\ref{fig:Ising_Plateau}).

\subsection{Strong coupling to the bath}
\label{sec:strongpert}

\begin{figure}[!h]
  \begin{center}
    \includegraphics[scale=0.34]{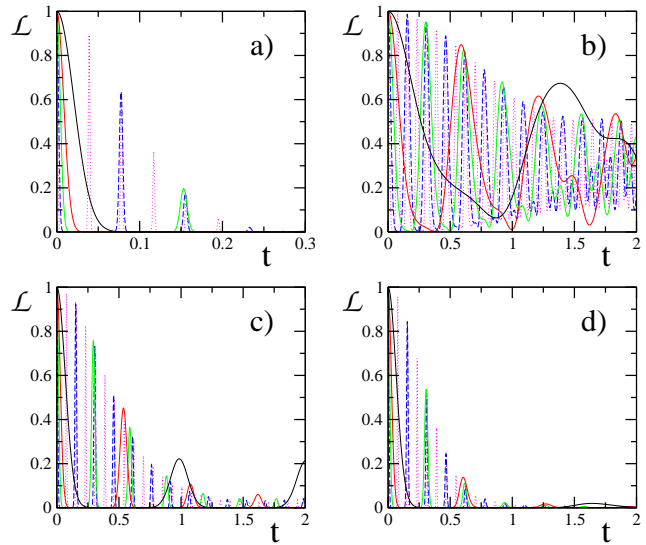}
    \caption{Loschmidt echo for the Ising model with $N=300$ spins as
      environment, $\lambda = 0.5$. The various panels are for different
      numbers of spins $m$ coupled to the qubit system
      and for various geometries:
      a) $m=300$ (central spin system);
      b) $m=3$, configuration B;
      c) $m=30$, configuration A;
      d) $m=30$, configuration B.
      Different curves correspond to various values of the coupling strength:
      $\epsilon=2$ (full black), 5 (full red), 10 (full green),
      20 (dashed blue), 40 (dotted magenta).}
    \label{fig:Ising_Univ}
  \end{center}
\end{figure}

Under certain conditions decoherence induced by the 
coupling with a bath manifests universal features: 
In Refs.~\cite{cucchietti2,zurek3} it has been shown that when the coupling
to the system drives a quantum phase transition in the environment,
the decay of coherences in the system is Gaussian in time,
with a width independent of the coupling strength.
In particular, for a central spin coupled to an $N$-spin Ising chain,
the Loschmidt echo in Eq.~\eqref{eq:loschquan} is characterized
by a Gaussian envelope modulating an oscillating term:
\beq
     \mathcal{L}(t)=\vert \cos (\epsilon t) \vert^{N/2} \: e^{- S_N^2 t^2} \, ,
\eeq
provided that
\beq
     \lambda < 1  \qquad {\rm and } \quad \lambda + \epsilon \gg 1 \, .
     \label{eq:strong}
\eeq
The oscillations are not universal, but the Gaussian width
$S_N^2$ depends only on the properties of the environment Hamiltonian,
in particular it is independent of the coupling strength $\epsilon$ 
and of the transverse magnetic field $\lambda$, while it is proportional to
the number $N$ of spins in the bath.
The case of the central spin model is illustrated
in Fig.~\ref{fig:Ising_Univ}a, where the different curves
stand for various values of $\epsilon$.

This universal behavior for strong couplings is present in more general
qubit-environment coupling, different from the central spin limit,
provided that the same conditions in Eq.~\eqref{eq:strong}
on $\lambda$ and $\epsilon$ hold.
In the Ising bath model, we checked that it remains valid also for
a number of system-bath links $m \neq N$ and for different geometries,
as it can be seen in Fig.~\ref{fig:Ising_Univ} (panels b-d).
The fast oscillations remain proportional to $\epsilon$, since they are
a consequence of the interaction $\propto \epsilon \, \tau_z \sigma^z_j$
between the qubit and the spins of the bath.
Instead in general the envelope is Gaussian only for small times,
but its width $S_N^2$ is independent of $\epsilon$.

\begin{figure}[!h]
  \begin{center}
    \includegraphics[scale=0.3]{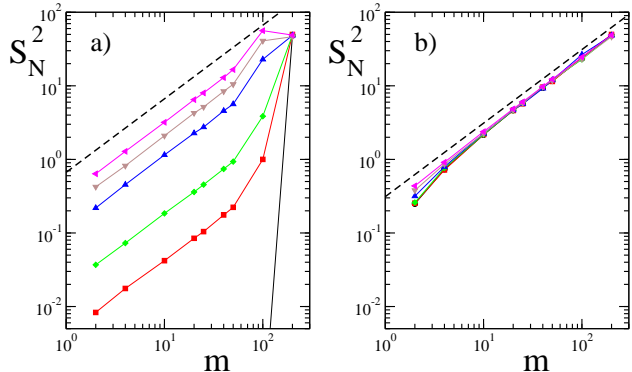}
    \caption{Width of the Gaussian envelope $S_N^2$ as a function of the number
      of links $m$ between the system and the bath, for different values of the
      transverse magnetic field: $\lambda = 0$ (circles), 0.1 (squares),
      0.2 (diamonds), 0.5 (triangles up), 0.7 (triangles down),
      0.9 (triangles left).
      The bath is modeled as an $N=200$ Ising chain with
      periodic boundary conditions;
      the coupling strength is kept fixed and equal to $\epsilon = 80$.
      The two panels correspond to: a) star-symmetric configuration (type A);
      b) non-symmetric configuration (type B).
      Dashed black curves represent a linear behavior $S_N^2 \sim m$
      and are plotted as guidelines.}
    \label{fig:Sigma_emme}
  \end{center}
\end{figure}

We found that, for the non-symmetric configuration B, $S_N^2$ depends only on
the number of links $m$, being proportional to it. No dependence from
the perturbation strength $\epsilon$, the size $N$ of the bath,
or the transverse magnetic field $\lambda$ has been observed, provided
that $m \gtrsim 10$, as it can be seen from Fig.~\ref{fig:Sigma_emme}b.
Indeed in the setup B the qubit behaves like in the central spin model:
in the regime of strong coupling, the effect of the environment spins
not directly linked to the qubit can be neglected
since they interact weakly, as compared to the system-bath interactions.
Therefore a system strongly coupled to $m$ neighbour spins
of an $N$-spin chain decoheres in the same way as if it was coupled
to all the spins of an $m$-spin chain.

\begin{figure}[!h]
  \begin{center}
    \includegraphics[scale=0.34]{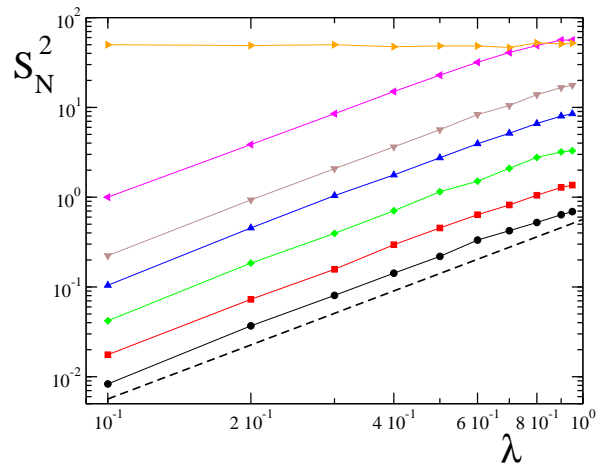}
    \caption{$S_N^2$ as a function of the transverse field $\lambda$
      in an $N=200$ spin Ising bath, for different
      numbers of system-bath links in the configuration A: $m=$ 2 (circles),
      4 (squares), 10 (diamonds), 25 (triangles up), 50 (triangles down),
      100 (triangles left), 200 (triangles right).
      The coupling strength is kept fixed and equal to $\epsilon = 80$.
      The case $m=200$ (triangles right) corresponds
      to the central spin model.}
    \label{fig:Conf_A_lambda}
  \end{center}
\end{figure}

The decoherence effects are richer in the geometry A.
In this setup the effect of the spins not directly coupled to the
qubit cannot be ignored, indeed we found that $S^2_N$ 
depends on the internal dynamics of the bath.
In particular we found that
\beq
     S_N^2 \sim m \, \lambda^2 \, ,
\eeq
where $\lambda$ is the transverse magnetic field
(see Fig.~\ref{fig:Sigma_emme}a and Fig.~\ref{fig:Conf_A_lambda}).
Notice that, as it can be seen from Fig.~\ref{fig:Sigma_emme}a, the case
$\lambda = 0$ is rather peculiar, since the system does not decohere at all,
unless it is uniformly coupled to all the spins of the environment.
This is due to the fact that an Ising chain in the absence of
a transverse field is not capable of transmitting quantum information:
in this system the density matrix of two non-neighbouring spins
evolves independently of the other spins.

\section{Conclusions}
\label{sec:conclusions}

In this work we analyzed a model of environment constituted by
an interacting one-dimensional quantum spin-1/2 chain.
We have shown that this type of baths can be experimentally engineered
in a fully controllable and tunable way by means of optical lattices,
and we provided a detailed description of the pulse sequence
needed to realize  a pure dephasing model for a qubit system
coupled to a spin bath.
The open quantum system model we proposed here is simpler to implement in an
experimental setup, with respect to the widely studied central spin model
in which the system is uniformly coupled to all the spins of the environment.
Nonetheless, we have analyzed the crossover from the single link to the
central spin model by taking in account different geometries and different
number of links, recovering, in the central spin limit, known results
present in literature, and characterizing new behaviors in between. 

The coherence loss in the system has been quantified  by means of the 
Loschmidt echo; the rich structure of the environment baths allowed us
to analyze the impact of the different phases and of the distance
from criticality, and more generally to study the connection between
entanglement inside the bath and decoherence effects in the system.

First we considered the case of one link
between the qubit and the chain.
We showed that at short times the Loschmidt echo decays as a Gaussian,
while for long times it approaches an asymptotic value which strongly
depends on the transverse field strength.
Criticality emerges in both short- and long-time behaviors:
the $XY$ spin bath is characterized by the first derivative
of the Gaussian width with respect to the field strength that diverges
logarithmically at $\lambda_c$, while the saturation values of the
Loschmidt echo at long times exhibits a sharp cusp in correspondence
to the critical point.
The decay rate at short times due to a Heisenberg spin bath is
maximum and constant throughout the whole critical region.
We also showed that, contrary to the central spin model,
if the environment is not critical, the decay of coherences
in the system is independent of the bath size.
Subsequently we related decoherence properties of the system
with quantum correlations inside the environment: we studied
the Gaussian decay rate at short times as a function of
the concurrence of two nearest neighbor spins in the bath.
We then analyzed a non-local system-bath coupling in which the
qubit is connected to the chain with multiple links, by considering
two different setups: a spin-symmetric and a non-symmetric one.
Finally we reported the existence of a coupling-independent regime,
with the only hypothesis of having a strong system-bath coupling.

\acknowledgments

We thank D. Cory, G. De Chiara and M. Rizzi for very useful discussions.
We acknowledge support from EC (grants QOQIP, RTNNANO, SCALA, SPINTRONICS
and EUROSQIP) and MIUR-PRIN. The present work has been partially
supported by the National Science Foundation through a grant for the Institute
for Theoretical Atomic, Molecular and Optical Physics at Harvard University
and Smithsonian Astrophysical Observatory, and it has been performed within the
``Quantum Information'' research program of Centro di Ricerca Matematica
``Ennio De Giorgi'' of Scuola Normale Superiore.

\appendix

\section{DMRG approach}
\label{DMRG}

An analytic solution for a spin coupled to a Heisenberg $XXZ$-model
(i.e. $\{ \Delta \neq 0$, $\lambda = 0 \}$ in Eq.~\eqref{eq:spinbath})
is not available, therefore a numerical approach is required.
We resort to the recently developed time-dependent
Density Matrix Renormalization Group (t-DMRG) with open boundary
conditions~\cite{white92,white04,schollwock,dmrgreview}), in order
to calculate the overlap between
the ground state $\ket{G}$ of $\Ham_g$ and the time evolution of the same
state under the Hamiltonian $\Ham_e$ (see Eq.~\eqref{eq:loschmidt}).

First, the static DMRG algorithm allows us to evaluate the ground state
$\ket{G}$ of $\Ham_g$. This algorithm in its first formulation~\cite{white92}
is an iterative numerical technique to find the ground state
of a one-dimensional system constituted by sites which possess
local and nearest-neighbor couplings, therefore it is well suited
for systems like the Hamiltonian in Eq.~\eqref{eq:spinbath}.
The key strategy of the DMRG is to construct a portion of the system,
called the system block, and then recursively enlarge it, until
the desired size is reached. At every step the basis of the corresponding
Hamiltonian is truncated, so that the size of the Hilbert space remains
manageable as the physical system grows. The truncation of the Hilbert
space is performed by retaining the eigenstates corresponding to the $D$
highest eigenvalues of the block's reduced density matrix.

The t-DMRG is then subsequently used, in order to simulate
the dynamics of $\ket{G}$ under the Hamiltonian $\Ham_e$.
The t-DMRG algorithm~\cite{white04} is an extension of the static DMRG, which 
follows the dynamics of a certain state $\ket{\psi}$ under the Hamiltonian
$\Ham$ of a nearest-neighbor one-dimensional system. The time evolution
can be implemented by using a second-order
Suzuki-Trotter decomposition~\cite{white04}
of the time evolution operator $U = e^{-i \Ham_e t}$:
\beq
     e^{-i \Ham_e t} \approx
     \left( e^{-i \mathcal{F} \frac{\delta t}{2}} e^{-i \mathcal{G} \delta t}
     e^{-i \mathcal{F} \frac{\delta t}{2}} \right)^n \, .
\eeq
Here we have first written the Hamiltonian as
$\Ham_e=\mathcal{F}+\mathcal{G}$ where
$\mathcal{F}$ contains the on-site terms and the even bonds while
$\mathcal{G}$ is formed by the odd bonds; $n=t/ \delta t$ is the time
expressed in number of trotter steps.
We also keep track of the state $\ket{G}$ in the new truncated basis
of $e^{-i \Ham_e t} \ket{G}$, so that we can straightforwardly evaluate
the overlap in Eq.~\eqref{eq:loschmidt}.
During the evolution the wave function is changing, therefore the
truncated basis chosen to represent the initial state
has to be updated by repeating the DMRG renormalization procedure
using the instantaneous state as the target state for the reduced
density matrix.

The results presented in this paper concerning Heisenberg spin baths,
have been obtained by performing time evolution within a second order
Trotter expansion, with a Trotter slicing $J \, \delta t = 10^{-3}$
and a truncated Hilbert space of dimension $D=100$.
The evaluation of the Loschmidt echo
with the t-DMRG is much more time- and memory-expensive than
the analytical approach of Eq.~\eqref{eq:det}, limiting the bath size
in our simulations up to $N \sim 10^2$ spins.

\section{Fermion correlation functions}
\label{fermion}

We provide an explicit expression for the two-point correlation matrix
$({\bf r})_{ij} = \langle \Psi^\dagger_i \Psi_j \rangle$ of the
operators ${\bf \Psi^\dagger} = \big( c_1^\dagger \ldots c_N^\dagger \,
c_1 \ldots c_N \big)$ on the ground state
of the system Hamiltonian $\Ham_g$.
The $2 N \times 2 N$ matrix {\bf r} is written in terms of the
Jordan Wigner fermions $\{ c_k, c_k^\dagger \}$ as:
\beq
     {\bf r} = \left( \begin{array}{cc}
     \big< c^\dagger_i c_j \big>_{i,j=1,N} &
     \big< c^\dagger_i c^\dagger_j \big>_{i,j=1,N} \\
     \big< c_i c_j \big>_{i,j=1,N} &
     \big< c_i c^\dagger_j \big>_{i,j=1,N}
     \end{array} \right) \, .
\eeq

Therefore it is sufficient to express
$\{ c_k, c_k^\dagger \}$ in terms of the normal mode operators
$\{ \eta_k^{(g)}, \eta_k^{(g) \dagger } \}$
which diagonalize the Hamiltonian $\Ham_g$, since
$< \eta_j^{(g)} \eta^{(g)}_k {}^\dagger > = \delta_{jk}$, the other
expectation values of the $\eta $'s on the ground state being zero.
By inverting Eq.~\eqref{eq:eta_c} we get
\beq
     c = {\bf g^T} \cdot \eta + {\bf h^T} \cdot \eta^\dagger \, ,
\eeq
from which it directly follows that
\beq
     {\bf r} = \left( \begin{array}{cc}
     {\bf h^{{ (g)} T}} \; {\bf h^{{(g)} }} &
     \quad
     {\bf h^{{(g)} T}} \; {\bf g^{{(g)} }} \\
     {\bf g^{{(g)} T}} \; {\bf h^{{(g)} }} &
     \quad
     {\bf g^{{(g)} T}} \; {\bf g^{{(g)} }}
     \end{array} \right). 
\eeq
The superscript $^{(g)}$ stands for the change-of basis-matrices
{\bf g} and {\bf h} relative to the Hamiltonian $\Ham_g$.

The last ingredient for the evaluation of the Loschmidt echo,
Eq.~\eqref{eq:det}, is the exponential $e^{i {\bf C} t}$.
We introduce the vector ${\bf \Gamma ^\dagger} =
\big( \eta_1^{(g)} {}^\dagger \ldots \eta^{(g)}_N {}^\dagger \:
\eta^{(g)}_1 \ldots \eta^{(g)}_N \big)$,
so that ${\bf \Psi} = {\bf U^\dagger} \, {\bf \Gamma}$, where
\beq
     {\bf U} = \left( \begin{array}{cc} {\bf g^{(e)}} & {\bf h^{(e)}} \\
     {\bf h^{(e)}} & {\bf g^{(e)}}
     \end{array} \right) \, .
\eeq
We can therefore rewrite Eq.~\eqref{eq:hame_c} as
\beq
     \Ham_e = \frac{1}{2} {\bf \Psi^\dagger \, C \, \Psi} =
     \frac{1}{2} {\bf \Gamma^\dagger \, U \, C \, U^\dagger \, \Gamma} \equiv
     \frac{1}{2} {\bf \Gamma^\dagger \, D \, \Gamma},
\eeq
where ${\bf D}$ is a $2 N \times 2 N$ diagonal matrix, whose elements 
are the energy eigenvalues of $\Ham_e$ and their opposites:
\beq
     {\bf D} = \left( \begin{array}{cc} {\bf E^{(e)}} &
     {\bf 0} \\ {\bf 0} & -{\bf E^{(e)}}
     \end{array} \right) \, .
\eeq
It then follows that ${\bf C = U^\dagger \, D \, U}$, from which
one can easily calculate the exponential $e^{i {\bf C} t}$.

\section{Central spin model}
\label{central}

If the qubit is uniformly coupled to all the spins of the chain,
the effect of the interaction in Eq.~\eqref{eq:hamint} is simply
of renormalizing the transverse field strength in the bath Hamiltonian:
for $\Delta = 0$ both Hamiltonians $\Ham_g$ and $\Ham_e$ correspond
to an $XY$-model with anisotropy $\gamma$, uniform couplings $J$ and
local magnetic field $\lambda$ and $\lambda + \epsilon$  respectively.
They can be diagonalized via a standard JWT,
followed by a Bogoliubov rotation~\cite{lieb61}.
The normal mode operators that diagonalize $\Ham_e$, satisfying
the fermion anti-commutation rules, are given by
\beq
     \eta_k^{(e)} = \sum_j \frac{e^{-2 \pi i j k/N}}{\sqrt{N}} \prod_{l < j}
     \sigma^x_l \left( u_k^{(e)} \sigma^+_j - i v_k^{(e)} \sigma^-_j \right) \, ,
\eeq
where the coefficients $u_k^{(e)} = \cos ( \theta_k /2 ), \:
v_k^{(e)} = \sin ( \theta_k /2 )$ depend on the angle
\beq
     \theta_k (\epsilon) = \arctan \left[ \frac{- \sin ( 2 \pi k /N)}
     {\cos (2 \pi k /N) - (\lambda + \epsilon)} \right] \, .
\eeq
The corresponding single quasi-excitation energy is given by
$E_k^{(e)} \equiv {\cal E}_k (\epsilon)$:
\beq
     {\cal E}_k (\epsilon) = 2 J \sqrt{\left[ \cos \Big( \frac{2 \pi k}{N} \Big)
     -  (\lambda + \epsilon ) \right]^2
     + \gamma^2 \sin^2 \Big( \frac{2 \pi k}{N} \Big)} \, .
     \label{eq:dispersion}
\eeq

The Hamiltonian $\Ham_g$ can be diagonalized in a similar way;
the corresponding normal-mode operators are connected
to the previous ones by a Bogoliubov transformation:
\beq
     \eta_{\pm k}^{(g)} = \cos (\alpha_k) \, \eta_{\pm k}^{(e)}
     -i \sin (\alpha_k) \, \eta_{\mp k}^{(e)} {}^\dagger \, ,
\eeq
with $\alpha_k = [\theta_k (0) - \theta_k (\epsilon)]/2$.

The spin chain is initially in the ground state $\ket{G}$
of $\Ham_g$; this state can be rewritten as a BCS-like state:
\beq
     \ket{G} = \prod_{k > 0} \left[ \cos (\alpha_k) - i \sin (\alpha_k)\,
     \eta^{(e)}_k {}^{\dagger}\, \eta^{(e)}_{-k} {}^{\dagger} \right] \ket{G_e}
\eeq
where $\ket{G_e}$ is the ground state of $\Ham_e$.
This expression allows one to rewrite the Loschmidt echo
$\mathcal{L} (t)$ in Eq.~\eqref{eq:loschmidt} in a
simple factorized form:
\beq \label{eq:loschquan}
     \mathcal{L}_{\scriptscriptstyle m=N}(t) = \prod_{k=1}^{N/2} \left[ 1 - \sin^2 (2 \alpha_k) \:
     \sin^2 ({\cal E}_k t) \right] \, .
\eeq

Eq.~\eqref{eq:loschquan} provides a straightforward formula in order
to calculate the Loschmidt echo for a central spin coupled uniformly
to all the spins of the bath. Nonetheless this model, although in some
circumstances it can reveal the emergence of a quantum phase transition
in the environment~\cite{quan,cucchietti2,ou06}, does not provide an effective
physical description of a standard reservoir, since the coupling with the
system is highly non-local and it can drive the evolution of the bath itself.
However it is useful throughout the paper, to test the convergence
of our results to the ones presented in~\cite{quan,cucchietti2,ou06},
in the limit $m \to N$.


\begin{thebibliography}{99}
\bibitem{revivals}
  A rough estimate of the revival times is given by $t^* = N/\bar{v}_\lambda$,
  where $\bar{v}_\lambda$ is the maximum phase velocity of the
  spin chain over all the normal mode velocities, in units of sites.
  Given the dispersion relation ${\cal E}_k$, then
  $\bar{v}_\lambda = \mathrm{Max}_{\bar{k}} (\partial_k {\cal E}) \vert_{\bar{k}}$.
  As an example, for the Ising chain (whose dispersion relation is given by
  Eq.~\eqref{eq:dispersion} with $\gamma = \epsilon = 0$) we have
  $\bar{v}_\lambda=2 \, \mathrm{min} (\lambda,1)$.
\end{thebibliography}
\end{document}